
\input harvmac
\baselineskip 12pt

\magnification\magstep1
\newdimen\itemindent \itemindent=32pt
\def\textindent#1{\parindent=\itemindent\let\par=\resetpar%
\indent\llap{#1\enspace}\ignorespaces}

\let\oldpar=\par
\def\resetpar{\oldpar\parindent=20pt\let\par=\oldpar}

\font\ninerm=cmr9 \font\ninesy=cmsy9
\font\eightrm=cmr8 \font\sixrm=cmr6
\font\eighti=cmmi8 \font\sixi=cmmi6
\font\eightsy=cmsy8 \font\sixsy=cmsy6
\font\eightbf=cmbx8 \font\sixbf=cmbx6
\font\eightit=cmti8
\def\eightpoint{\def\rm{\fam0\eightrm}
  \textfont0=\eightrm \scriptfont0=\sixrm \scriptscriptfont0=\fiverm
  \textfont1=\eighti  \scriptfont1=\sixi  \scriptscriptfont1=\fivei
  \textfont2=\eightsy \scriptfont2=\sixsy \scriptscriptfont2=\fivesy
  \textfont3=\tenex   \scriptfont3=\tenex \scriptscriptfont3=\tenex
  \textfont\itfam=\eightit  \def\it{\fam\itfam\eightit}%
  \textfont\bffam=\eightbf  \scriptfont\bffam=\sixbf
  \scriptscriptfont\bffam=\fivebf  \def\bf{\fam\bffam\eightbf}%
  \normalbaselineskip=9pt
  \setbox\strutbox=\hbox{\vrule height7pt depth2pt width0pt}%
  \let\big=\eightbig  \normalbaselines\rm}
\catcode`@=11 %
\def\eightbig#1{{\hbox{$\textfont0=\ninerm\textfont2=\ninesy
  \left#1\vbox to6.5pt{}\right.\n@space$}}}
\def\vfootnote#1{\insert\footins\bgroup\eightpoint
  \interlinepenalty=\interfootnotelinepenalty
  \splittopskip=\ht\strutbox %
  \splitmaxdepth=\dp\strutbox %
  \leftskip=0pt \rightskip=0pt \spaceskip=0pt \xspaceskip=0pt
  \textindent{#1}\footstrut\futurelet\next\fo@t}
\catcode`@=12 %

\input amssym.def
\input amssym
\def \bR{{\Bbb R}}
\font \bigbf=cmbx10 scaled \magstep1

\def \l{\langle}
\def \r{\rangle}
\def \de{\delta}
\def \si{\sigma}
\def \Ga{\Gamma}

\def \ga{\gamma}
\def \nab{\nabla}

\def \pr{\partial}
\def \hU{{\hat U}}
\def \hV{{\hat V}}

\def \hx{{\hat x}}
\def \hr{{\hat r}}

\def \hmu{{\hat \mu}}
\def \d{{\rm d}}

\def \tr{{\rm tr \, }}
\def \bO{\bar{\cal {O}}}
\def \rO{{\rm O}}
\def \vep{\varepsilon}
\def \half{{\textstyle {1 \over 2}}}

\def \quar{{\textstyle {1 \over 4}}}

\def \hh{{1\over 2}}
\def \ts{ \textstyle}

\def \A{{\cal A}}
\def \B{{\cal B}}
\def \C{{\cal C}}
\def \D{{\cal D}}
\def \E{{\cal E}}

\def \G{{\cal G}}
\def \X{{\cal X}}
\def \I{{\cal I}}
\def \O{{\cal O}}

\def \R{{\cal R}}
\def \L{{\cal L}}

\def \ep{\epsilon}

\lref\one{H. Osborn and A. Petkou, Ann. Phys. (N.Y.) 231 (1994) 311.}
\lref\zam{A.B. Zamolodchikov, {\ JETP Lett.} {43} (1986) 43; 
{Sov. J. Nucl. Phys.} { 46} (1988) 1090\semi
A.W.W. Ludwig and J.L. Cardy, {Nucl. Phys.} {B285} [FS19] (1987) 687.}
\lref\irfp{ D.J. Gross, {\it in} `Methods in Field Theory', Les Houches Lectures
1975, ed. R. Balian and J. Zinn-Justin  (North-Holland, 1976)\semi
P. Olesen, {Nucl. Phys.} {B104} (1976) 125\semi
T. Banks and A. Zaks, {Nucl. Phys.} {B196} (1982) 189.}
\lref\jack{I. Jack and H. Osborn, {Nucl. Phys.} {B343} (1990) 647.}
\lref\cth{J.L. Cardy, {Phys. Lett.} {B215} (1988) 749\semi
H. Osborn, {Phys. Lett.} {B222} (1989) 97\semi
G.M. Shore, {Phys. Lett.} {B253} (1991) 380; {B256} (1991) 407.}
\lref\friedan{A. Cappelli, D. Friedan and J.I. Latorre, {Nucl. Phys.}
{B352} (1991) 616.}
\lref\cap{A. Cappelli, J.I. Latorre and X. Vilas\'{\i}s, {Nucl. Phys.} {B376}
(1992) 510.}
\lref\deser{S. Deser and A. Schwimmer, {Phys. Lett.} {B309} (1993), 279.}
\lref\riegert{R.J. Riegert, {Phys. Lett.} {134B} (1984) 56.}
\lref\barv{A.O. Barvinsky and G.A. Vilkovisky, {Nucl. Phys.} {B282} (1987) 163;
{Nucl. Phys.} {B333} (1990) 471, 572\semi
A.O. Barvinsky, Yu. V. Gusev, V.V. Zhytnikov and G.A. Vilkovisky,
``Covariant Perturbation Theory (IV). Third Order in the Curvature",
University of Manitoba preprint;
{J. Math.  Phys.} {35} (1994) 3525, 3543; {Nucl. Phys.} {B439} (1995) 561;
{Class. Quantum Grav.} {12} (1995) 2157\semi
A.O. Barvinsky, A.G. Mirzabekian and  V.V. Zhytnikov, gr-qc/9510037\semi
A.G. Mirzabekian, G.A. Vilkovisky and V.V. Zhytnikov, {Phys. Lett.}
{B369} (1996) 215, hep-th/9510205.}
\lref\west{M. Sohnius and P. West, {Phys. Lett.} {100B} (1981) 245.}
\lref\howe{P. Howe, M. Sohnius and P. West, {Phys. Lett.} {124B} (1983) 55.}
\lref\duff{M.J. Duff, Class. Quantum Grav. {11} (1994) 1387.}
\lref\cft{P. Ginsparg and also J. Cardy, {\it in} `Champs, Cordes et 
Ph\'enom\`enes
Critiques', (E. Br\'ezin and J. Zinn-Justin eds.) North Holland, Amsterdam
1989.}
\lref\cardy{J. Cardy, {Nucl. Phys.} {B290 [FS20]} (1987) 355.}
\lref\pet{A.C. Petkou, Ann. Phys (N.Y.), 249 (1996) 180, hep-th/9410093.}
\lref\diff{D.Z. Freedman, K. Johnson and J.I. Latorre, {Nucl. Phys.}
{B371} (1992) 353\semi
D.Z. Freedman, G. Grignani, K. Johnson and N. Rius, {Ann. Phys. (N.Y.)}
{218} (1992) 75.}
\lref\form{J.A.M. Vermaseren, Symbolic Manipulation with FORM, version 2.
Computer Algebra Nederland, Amsterdam 1990.}
\lref\proc{H. Osborn, DAMTP preprint 93/67, {\it in} Proceedings of the XXVII
Ahrenshoop International Symposium, DESY 94-053.}
\lref\fixp{N. Seiberg, {Nucl. Phys.} {B435} (1995) 129, hep-th/9411149\semi
R.G. Leigh and M.J. Strassler, {Nucl. Phys.} {B447} (1995) 95, 
hep-th/9503121; Phys.Lett. B352 (1995) 20, hep-th/9505088\semi
D. Kutasov, Phys. Lett. B351 (1995) 230, hep-th/9503086\semi
D. Kutasov and A. Schwimmer, Phys. Lett. B354 (1995) 315, hep-th/9505004\semi
K. Intriligator, Nucl. Phys. B448 (1995) 187, hep-th/9505051\semi
K. Intriligator, R.G. Leigh and M.J. Strassler, hep-th/9506148\semi
I.I. Kogan, M. Shifman and A. Vainshtein, Phys. Rev. {D53} (1996) 4526,
hep-th/9507170.}
\lref\scft{P.C. Argyres and M.R. Douglas,  Nucl. Phys. {B448} (1995) 93,
hep-th/9505062\semi
P.C. Argyres, M.R. Plesser, N. Seiberg and E. Witten, Nucl. Phys.
{B461} (1996) 71, hep-th/9511154\semi
T. Eguchi, K. Hori, K. Ito and S-K. Yang, {Nucl. Phys.} {B471} (1996) 430,
hep-th/9603002.}
\lref\gris{D. Anselmi, M. Grisaru and A. Johansen, preprint HUTP-95/A048,
BRX-TH-388, hep-th/9601023.}
\lref\hwest{P. Howe and P West, preprint KCL-TH-95-9, hep-th/9509140.}
\lref\scth{F. Bastianelli, Phys. Lett. {B369} (1996) 249, hep-th/9510037.}
\lref\free{D.Z. Freedman, MIT preprint CTP-2020, {\it in} Proceedings of
the Stony Brook Conference on Strings and Symmetries, 1991.}
\lref\schreier{E.J. Schreier, Phys. Rev. {D3} (1971) 980.}
\lref\crewther{R.J. Crewther, Phys. Rev. Lett. 28 (1972) 1421.}
\lref\castro{A.H. Castro Neto and E. Fradkin, {Nucl. Phys.} {B400} (1993) 525.}
\lref\salam{G. Mack and A. Salam, Ann. Phys. (N.Y.) {53} (1969) 174.}
\lref\curve{I. Jack and H. Osborn, {Nucl. Phys.} {B234} (1984) 331.}
\lref\poly{A.M. Polyakov, {Phys. Lett.} {B103} (1981) 207.}
\lref\vasilev{A.N. Vasil'ev, Yu.M. Pis'mak and Yu.R. Khonkonen, {Theoretical
and Mathematical Physics} {46} (1981) 104; {47} (1981) 465;
{50} (1982) 127\semi
A.N. Vasil'ev and M.Yu. Nalimov, {Theoretical
and Mathematical Physics} {55} (1983) 423; {56} (1984) 643\semi
A.N. Vasil'ev, M.M. Perekalin and Yu.M. Pis'mak, {Theoretical
and Mathematical Physics} {55} (1983) 529; {60} (1985) 846.}
\lref\lang{K. Lang and W. R\"uhl, {Zeit f. Phys. C} {50} (1991) 285; {51} (1991)
127; {Nucl. Phys.} {B377} (1992) 371; {B400}[FS] (1992) 597;
{B402} (1993) 573; {Zeit f. Phys. C} {61} (1994) 495; {63} (1994) 531\semi
A.N. Vasil'ev, and A.S. Stepanenko, {Theoretical and Mathematical Physics} 
{95} (1993) 160.}
\lref\moore{A. Losev, G. Moore, N. Nekrasov and S. Shatashvili, preprint
PUPT-1564, hep-th/9509151.}
\lref\odintsov{E. Elizalde, A.G. Jacksenaev, S.D. Odintsov and I.L. Shapiro,
{Class. Quantum Grav.} 12 (1995) 1385.}
\lref\haw{S.W. Hawking and G.F.R. Ellis, The Large Scale Structure of
Space-time, Cambridge University Press, Cambridge 1973.}
\lref\pan{S. Paneitz, MIT preprint (1983)\semi
M. Eastwood and M. Singer, Phys. Lett. {107A} (1985) 73.}
\lref\branson{T.P. Branson, Math. Scand. 57 (1985) 293\semi
V. W\"unsch, Math. Nachr. 129 (1986) 269.}
\parskip 5pt
{\nopagenumbers
\rightline{DAMTP/96-7}
\rightline{hep-th/9605009}
\vskip 2truecm
\centerline {\bigbf Conserved Currents and the Energy Momentum Tensor in}
\vskip 5pt
\centerline {\bigbf Conformally Invariant Theories for General Dimensions}
\vskip 2.0 true cm
\centerline {J. Erdmenger and H. Osborn\footnote{*}{email: 
je10001@damtp.cam.ac.uk and ho@damtp.cam.ac.uk}}
\vskip 10pt
\centerline {Department of Applied Mathematics and Theoretical Physics,}
\centerline {Silver Street, Cambridge, CB3 9EW, England}
\vskip 2.0 true cm
\font \abs=cmr10 at 10 true pt
\font \absit=cmti10 at 10 true pt
{\abs \openup -1\jot 
\parindent 1.5cm{

{\narrower\smallskip\parindent 0pt
The implications of conformal invariance, as relevant in quantum field theories
at a renormalisation group fixed point, are analysed with particular reference
to results for correlation functions involving conserved currents and the 
energy momentum tensor. Ward identities resulting from conformal invariance 
are discussed. Explicit expressions for two and three point functions,
which are essentially determined by conformal invariance, are obtained. As
special cases we consider the three point functions for two vector and an
axial current in four dimensions, which realises the usual anomaly simply
and unambiguously, and also for the energy momentum tensor in general dimension
{\absit d}.
The latter is shown to have two linearly independent forms in which the 
Ward identities are realised trivially, except if {\absit d} = 4,
when the two forms become degenerate.
This is necessary in order to accommodate the two independent forms present
in the trace of the energy momentum tensor on curved space backgrounds for
conformal field theories in four dimensions. The coefficients of the two
trace anomaly terms are related to the three parameters describing the
general energy momentum tensor three point function. The connections with
gravitational effective actions depending on a background metric are
described. A particular form due to Riegert is shown to be unacceptable.
Conformally invariant expressions for the effective action in four dimensions
are obtained using the Green function for a differential operator which has
simple properties under local rescalings of the metric.

\narrower}}
\openup 1\jot}
\vfill\eject}
\pageno=1

\leftline{\bigbf 1 Introduction}
\medskip
Conformally invariant field theories are well known and have been  much 
studied in two dimensions \cft. Recently the demonstration of existence
of fixed points in a large class of $N=1$ and $N=2$ supersymmetric theories
at which conformal invariance holds \refs{\fixp,\scft}, as well as the
long standing cases of finite $N=4$ and $N=2$ theories \refs{\west,\howe}
in which the $\beta$-function vanishes identically, has revitalised the
interest in studying non trivial conformal theories in four dimensions
\refs{\gris,\hwest}. Such infra-red fixed points arise in asymptotically 
free supersymmetric theories with a single gauge coupling $g$ but the presence
of additional fields besides the gauge vector supermultiplet generates
the zero of the $\beta(g)$ which is necessary for a fixed point. The essential
mechanism is basically identical to that which gives a $\beta$-function zero
at two loops in ordinary gauge field theories with suitably adjusted numbers
of fermions \refs{\irfp,\jack}
(the use of perturbative results may be justified for large $N_c$ for 
gauge group $SU(N_c)$ with the numbers of fermions in the fundamental
representation $N_f = \rO (N_c)$).

Since conformal invariance provides very non trivial constraints it is 
possible to hope \moore\ that exact results for four dimensional theories
may also be
obtained. In any event using conformal invariance allows for significantly
simplified calculations of the scale dimensions of operators at the fixed point
than would be allowed in conventional perturbative approaches based on 
expansions about free theories \refs{\vasilev,\lang}.
One of the crucial consequences of conformal
invariance even in dimensions $d>2$ is that the functional forms of
two and three point functions of operators are essentially determined
with no arbitrary functions present. For operators with spin there may
however be two or more linearly independent forms compatible with conformal
invariance which are possible for the three point functions \one.

In this paper we extend recent investigations by one of us \one\
to discuss in 
detail the form of the three point functions involving conserved vector
currents $V_\mu$ and the energy momentum tensor $T_{\mu\nu}$ in general
$d$ Euclidean dimensions. This is motivated by analogy with two dimensional
conformal field theories where the parameters which occur in the two and
three point functions of these operators, such as the Virasoro central
charge $c$, play a vital role in specifying the theory. If any result like
the Zamolodchikov $C$-theorem \zam\ is to hold in four space time dimensions
\refs{\jack,\cth,\friedan,\castro,\scth}
then it must involve quantities which are well defined in the conformal
limit \cap. One suggestion for a possible generalisation involves the
coefficient of the term in the
trace of the energy momentum tensor on a curved space background involving
the topological Euler density \cth\ (in two dimensions this is proportional to
$c$). As shown later, in four dimensions this parameter is connected directly
with one of three linearly independent forms for the conformal invariant
three point function of the energy momentum tensor.

The conserved current and energy momentum tensor of course satisfy the
equations
$$
\pr_\mu V_\mu = 0 \, , \qquad \pr_\mu T_{\mu\nu} = 0 \, , \quad 
T_{\mu\nu} = T_{\nu\mu} \, , \quad T_{\mu \mu}=0 \, ,
\eqno (1.1)
$$
where conformal invariance dictates that $V_\mu$ and $T_{\mu\nu}$ must have
scale dimension $d-1$ and $d$ respectively. In our discussion it is
important to recognise that the operators $V_\mu$ and $T_{\mu\nu}$ are not
uniquely defined. It is possible to add to $V_\mu$ or $T_{\mu\nu}$ terms
which lead to the same conserved charge or generators of the conformal group.
For the current $V_\mu$ the arbitrariness has the form
$$
V_\mu \sim V_\mu + \pr_\nu F_{\mu\nu} \, , \quad  F_{\mu\nu} = - 
F_{\nu\mu} \ \hbox{or} \ F_{\mu\nu} = F_{[\mu\nu]} \, ,
\eqno (1.2) $$
while for the energy momentum tensor correspondingly
$$
T_{\mu\nu}  \sim T_{\mu\nu} + \pr_\si \pr_\rho C_{\mu\si\rho\nu} \, ,
\eqno (1.3) $$
where $C_{\mu\si\rho\nu}$ has the symmetries of the Weyl tensor
$$
C_{\mu\si\rho\nu} =  C_{[\mu\si][\rho\nu]} \, , \quad  C_{\mu[\si\rho\nu]}
= 0 \, , \quad  C_{\mu\si\rho\mu} = 0 \, ,
\eqno (1.4) $$
with $F_{\mu\nu}$ and $C_{\mu\si\rho\nu}$ each of dimension $d-2$. The
additional terms in (1.2) and (1.3) automatically satisfy (1.1). It is also 
crucial that both operators are quasi-primary operators which means that 
they transform homogeneously as tensor operators 
under conformal transformations. In particular theories operators 
with the assumed properties of $F_{\mu\nu}$ and $C_{\mu\si\rho\nu}$ need
not exist (for $d=2$ or $3$ $C_{\mu\si\rho\nu}$ is necessarily zero).
Nevertheless the freedom exhibited in (1.2) or (1.3) with (1.4)
is ultimately behind the existence of more than one linearly independent
form for the three point functions involving $V_\mu$ or $T_{\mu\nu}$. In some
cases, where there are no
non trivial Ward identities, the three point functions can be written
just as if the current or energy momentum tensor was given by the terms
involving $F$ or $C$ in (1.2) or (1.3). When this is feasible this ensures
that the resulting expression is less singular since, in the spirit of
differential regularisation \diff, derivatives have been pulled out.

In discussing two or three point functions of conserved currents it is
essential to pay careful attention to anomalies in the various Ward identities.
Such anomalies are most succinctly expressed in the presence of background
fields. Thus for the fermion axial current with a gauge field $A_\mu$ coupled
to the vector current the well known anomaly in four dimensions, if gauge
invariance with respect to gauge transformations on $A_\mu$ is preserved, has
the form
$$
\pr_\mu \l {\bar \psi} \gamma_\mu \gamma_5 \psi \r_A = {1\over 16\pi^2}
\, \ep_{\mu\nu\si\rho} F_{\mu\nu}F_{\si\rho} \, , \quad
F_{\mu\nu} = \pr_\mu A_\nu - \pr_\nu A_\mu \, .
\eqno (1.5) $$
For the energy momentum tensor it is natural to consider the theory 
extended to a general curved space with metric $g_{\mu\nu}$ so that the
vacuum energy functional, or effective action, is a functional $W(g,A)$
and we may define the energy momentum tensor and conserved current in this
background by
$\sqrt g \l T_{\mu\nu} \r_{g,A} = - 2\de W / \de g^{\mu\nu}$ and
$\sqrt g \l V^\mu \r_{g,A} = - \de W / \de A_\mu$. In this case the
gauge symmetry (when $\de A_\mu = \pr_\mu \lambda$) and invariance under
diffeomorphisms (when $\de g^{\mu\nu} = \L_v g^{\mu\nu} = -\nab^\mu v^\nu
- \nab^\nu v^\mu$ and $\de A_\mu = \L_v A_\mu = 
v^\nu\pr_\nu A_\mu + \pr_\mu v^\nu A_\nu$ for $\L_v$ the Lie derivative),
which are assumed to be preserved in the quantum theory, lead to
$$
\nab_\mu \l V^\mu \r_{g,A} = 0 \, , \quad
\nab^\mu \l T_{\mu\nu} \r_{g,A} + F_{\nu\mu} \l V^\mu \r_{g,A} = 0 \, .
\eqno (1.6) $$
For a theory which is conformal on flat space there is now an anomaly in
the energy momentum tensor trace \duff\ $\!$\footnote{*}{This review 
contains a comprehensive list of references on the trace anomaly.}
which may be assumed to be of the form in four dimensions  
$$ g^{\mu\nu}\l T_{\mu\nu} \r_{g,A} =  -  \kappa \, \quar
F^{\mu\nu} F_{\mu\nu} - \beta_a F - \beta_b G  \, ,
\eqno (1.7) $$
where
$$
F = C^{\alpha\beta\gamma\de} C_{\alpha\beta\gamma\de} \, , \quad
G = \quar \ep^{\mu\nu\si\rho}\ep_{\alpha\beta\gamma\beta}
R^{\alpha\beta}{}_{\mu\nu} R^{\gamma\delta}{}_{\si\rho} \, ,
\eqno (1.8) $$
with $C_{\alpha\beta\gamma\de}$ the Weyl tensor which is given in terms of
the Riemann tensor in (A.3). $G$ is the topological Euler density in four
dimensions. The curved space anomaly implies an anomaly of
the three point function when restricted to flat space \refs{\one,\proc},
$$ \eqalign { \!\!\!\!\!\!\!\!\!\!\!\!
\l T_{\mu\mu} (x)  \, T_{\si\rho} (y) \, T_{\alpha \beta} (z) \r &{}
= 2  \bigl (\de^4(x-y)+\de^4(x-z)\bigl)\,
\l T_{\si\rho} (y) \, T_{\alpha \beta} (z) \r \cr
& \!\!\!\!\! \, - 32 \beta_a\,
\E^C{}_{\!\! \si\ep\eta\rho,\alpha\gamma\de\beta}\, \pr_{\ep} \pr_{\eta} 
\de^4(x-y) \pr_{\gamma} \pr_{\de} \de^4(x-z) \cr
&  \!\!\!\!\! \, + 4 \beta_b \bigl \{
\ep_{\si\alpha\ep\kappa}\ep_{\rho\beta\eta\lambda}\pr_\kappa
\pr_\lambda \bigl( \pr_\ep \de^4 (x-y) \pr_\eta \de^4 (x-z) \bigl ) {}
+ \si \leftrightarrow \rho \bigl \} ,\cr }
\eqno (1.9) $$
where $\E^C$ denotes the projection operator onto tensors with the symmetries
(1.4) and is given explicitly in (A.1) for general $d$, it may be
defined by $\E^C{}_{\!\! \mu\si\rho\nu,\alpha\gamma\delta\beta} =\pr
C_{\mu\si\rho\nu}/\pr C_{\alpha\gamma\delta\beta}$.

As a precursor to describing various applications of conformal invariance
we review in the next section
the general form of conformal transformations when $d>2$ and
define quasi-primary fields by their conformal transformation properties.
The statements made above concerning (1.2) and (1.3) are justified
and the ingredients necessary for a general construction of two and three
point functions introduced. The consequences of conformal invariance for
Ward identities involving the energy momentum tensor are also briefly
described. The general results are first applied in section 3
to the three point function of two vector currents and an axial current.
Using conformal invariance we obtain an expression in which vector current
conservation is automatic and the divergence of the axial current give
rise unambiguously to the standard expression for the axial anomaly. Of
course this result is a necessary consistency test but our approach may be
regarded as allowing a simple derivation of the anomaly which is in the
framework of differential
regularisation. In section 4 we apply the same techniques to the three
point function of two vector currents and the energy momentum tensor.
For this example we can effectively write $V_\mu = \pr_\nu F_{\mu\nu}$ which
allows less singular expressions to be obtained since the dimension of
$F_{\mu\nu}$ is $d-2$ rather than $d-1$ for $V_\mu$. In four dimensions
the results are compatible with the trace anomaly for external background gauge
fields. In section 5 we recover the previous result \one\ that there are only
three linearly independent forms for the three point function 
$\l T_{\mu \nu}(x) \,T_{\si\rho} (y) \, T_{\alpha \beta}  (z) \r$. The
expressions obtained are simplified by using expressions in which
the symmetry is manifest so that it is only necessary to impose the
conservation equation. In section 6 it is shown how to obtain
less singular expressions making use of (1.3) although this only allows for
two linearly independent expressions. The results trivially 
satisfy the Ward identities
for the divergence and trace of $T_{\mu\nu}$. In the latter case when
$d=4$ there is an anomaly of the expected form corresponding to the trace
of the energy momentum tensor on a curved space background. For the
anomaly associated with the Euler density $G$ the mechanism is in accord 
with the suggestions of Deser and Schwimmer \deser\ since it involves tensors
which vanish identically when $d=4$ but which are combined with a pole in
$\vep = 4-d$. We derive an  expression for the coefficient of the
anomaly in terms of the parameters of the general three point function
which is compatible with results obtained by explicit calculation
for free fields. In section 7 we also discuss the form of the gravitational
effective action which may be expressed as a non local function of the
curvature and which is directly connected with the energy momentum tensor
correlation functions on flat space. We show that a particular elegant
expression due to Riegert \riegert, which generates the $G$ term in the
trace anomaly in (1.7), is incompatible with conformal invariance for the
corresponding expression for the energy momentum tensor three point function
on flat space since $\l T_{\mu\nu}(x) \r_g$ does not fall off sufficiently
rapidly in this case for large $|x|$ for asymptotically flat spaces.
Other forms for parts of the effective action $W$ in four dimensions which are
compatible with conformal invariance are also considered. To achieve this
we introduce a second order differential operator $\Delta^F$ which acts on
2-forms, or antisymmetric tensors, which has nice properties under local
rescalings of the metric, similar to the operator $-\nab^2 + {1\over 6}R$
acting on scalars. Effective actions constructed using the Green function
for $\Delta^F$ are then in accord with our conformal invariance results
when reduced to flat space. In a conclusion we show in general how
in the three point function of a conserved vector current or the energy
momentum tensor with operators of different dimension the we can always
write it in a form so that effectively $V_\mu = \pr_\nu F_{\mu\nu}$ or
$T_{\mu\nu} = \pr_\si \pr_\rho C_{\mu\si\rho\nu}$.
Some lengthy mathematical formulae are for convenience relegated to
three appendices.

The results of this paper are a development of earlier work by one of us
\one\ but we endeavour to simplify the presentation and make more systematic
use of (1.2,3) and also consider the implications for the effective action
on curved space.
\bigskip
\leftline{\bigbf 2 Conformal Invariance}
\medskip

The group of conformal transformations  acting on $\bR^d$
is defined by coordinate transformations
such that
$$ x_\mu \to x'{}_{\! \mu} (x) \, ,
\quad dx'{}_{\! \mu} dx'{}_{\! \mu} = \Omega(x)^{-2} dx_\mu dx_\mu \ .
\eqno (2.1) $$
For an infinitesimal transformation we may write
$$ \eqalign {
x'{}_{\! \mu} (x) = x_\mu + v_\mu (x) , \qquad \Omega (x) = 1 -
\si_v (x) \, , \cr
\quad \pr_\mu v_\nu + \pr_\nu v_\mu =  2 \si_v
\de_{\mu\nu} \, , \qquad \si_v = {1\over d}\, \pr {\cdot v} \, . \cr}
\eqno (2.2) $$
Except for $d=2$ the general solution of (2.2) has the form
$$ v_\mu (x) = a_\mu + \omega_{\mu\nu} x_\nu + \lambda x_\mu +
b_\mu x^2 - 2 x_\mu b {\cdot x} \, , \quad  \omega_{\mu\nu} = -
 \omega_{\nu\mu} \, , \quad \si_v(x) = \lambda - 2 b {\cdot x} \, ,
\eqno (2.3) $$
representing infinitesimal translations, rotations, scale transformations
and special conformal transformations. 
For any such conformal transformation we may define a local orthogonal
transformation by
$$\R_{\mu \alpha}(x) = \Omega (x)
{\pr x'{}_{\! \mu} \over \pr x_\alpha} \ , \quad \R_{\mu \alpha} (x)
\R_{\nu \alpha} (x) = \delta_{\mu \nu} \, ,
\eqno (2.4) $$
which in $d$ dimensions is a matrix belonging to $O(d)$. Apart from
conformal transformations which are connected to the identity there
are also inversions. For an inversion through the origin this has the form
$$ x'{}_{\!\mu} = {x_\mu \over x^2} \, ,
\quad \Omega (x) = x^2 \, , \quad
\R_{\mu \nu}(x) = I_{\mu \nu}(x) 
= \de_{\mu \nu} - 2 \, {x_\mu x_\nu \over x^2} \, .
\eqno (2.5) $$
Any conformal transformation can be generated by combining inversions with
rotations and translations. In $d$ dimensions the conformal group is
isomorphic with $O(d+1,1)$.

The construction of conformally invariant forms for correlation functions
of quasi-primary fields depend on recognising that for arbitrary
conformal transformations the inversion matrix $I$,
as defined in (2.5), plays the role of a parallel transport since for
two points $x,y$ we have
$$
I_{\mu \nu}(x'-y') = \R_{\mu \alpha}(x) \R_{\nu \beta}(y) I_{\alpha\beta}
(x-y) \, , \quad (x'-y')^2 = {(x-y)^2\over \Omega(x) \Omega(y)}\, .
\eqno (2.6) $$
Furthermore for three points $x,y,z$ we may define vectors at each point
which transform homogeneously under conformal transformation. At $z$ the
associated vector is given by
$$
Z_\mu = \half \pr^z {}_{\! \mu} \ln {(z-y)^2\over (z-x)^2} = 
{(x-z)_\mu \over (x-z)^2} - {(y-z)_\mu \over (y-z)^2} \, , \quad
Z^2 = {(x-y)^2 \over (x-z)^2 (y-z)^2} \, ,
\eqno (2.7) $$
so that 
$$
Z'{}_{\! \mu} = \Omega(z) \R_{\mu \alpha}(z) Z_\alpha \, .
\eqno (2.8) $$
$X_\mu$ and $Y_\mu$, which are vectors at $x$ and $y$, are defined by cyclic
permutation. Conformal invariance requires various identities, important
ones for subsequent use are
$$
I_{\mu\alpha}(x-z) Z_\alpha = - {(x-y)^2\over (z-y)^2} \, X_\mu \, , \quad
I_{\mu\alpha}(x-z)I_{\alpha\nu}(z-y) = I_{\mu \nu}(x-y) + 2 (x-y)^2
X_\mu Y_\nu \, ,
\eqno (2.9) $$
together with obvious permutations (note that for $x\leftrightarrow y$
$Z\to -Z$ while $ X \leftrightarrow - Y$).

A quasi-primary field $\O^i(x)$, where $i$ denotes components in some space
on which a representation of $O(d)$ acts,
is defined by the transformation properties \salam
$$ \O^i(x) \to \O'{}^i(x') = \Omega(x)^{\eta}  D^i {}_{\! j}(\R (x) )
\O^j (x) \, ,  
\eqno (2.10) $$
where $\eta$ is the scale dimension and $D^i {}_{\! j}(R)$ denotes the
representation for $R_{\mu\nu}\in O(d)$. This result reduces to the standard
transformation rules for translations and rotations. For an infinitesimal
transformation as in (2.2)
$$
\de_v \O^i(x) = -(L_v \O)^i(x) \, , \quad
L_v = v{\cdot \pr} + \eta \si_v - \half \pr_{[\mu} v_{\nu]} \, s_{\mu\nu} \, ,
\eqno (2.11) $$
where $(s_{\mu\nu})^i{}_{\! j} = - (s_{\nu\mu})^i{}_{\! j}$
are the generators of $O(d)$ for the representation which acts on $\O$.
It is easy to verify that $L_v$ obey the required  Lie algebra 
$[L_v , L_{v'}]=L_{[v,v']}$ for the conformal group.

It is clear from the definition in (2.10) that the derivative of a
quasi-primary field, $\pr_\mu\O$, is in general no longer quasi-primary, since
there are additional inhomogeneous terms involving $\O$ in the transformation.
In special cases these terms may cancel. For a vector
field $V_\mu$ of dimension $\eta$ the application of (2.10) for an 
infinitesimal conformal transformation gives
$$
\de_v V_\mu = - L_v V_\mu = - \big ( v{\cdot \pr} + \eta \si_v \big ) V_\mu 
- \pr_{[\mu} v_{\nu]} \, V_\nu \, ,
\eqno (2.12) $$
since for this representation $(s_{\mu\nu})_{\alpha\beta} =
-\de_{\mu\alpha}\de_{\nu\beta}+\de_{\mu\beta}\de_{\nu\alpha}$.
It is then easy to see using (2.3), so that $ \pr_{[\mu} v_{\nu]} = -
\omega_{\mu\nu} + 2 (x_\mu b_\nu - x_\nu b_\mu )$,
$$
\pr_\mu \de_v V_\mu =  - \big ( v{\cdot \pr} + (\eta + 1) \big ) \pr_\mu V_\mu 
+ 2 (\eta -d + 1 ) b_\mu V_\mu \, .
\eqno (2.13) $$
Hence if $\pr_\mu V_\mu$ is to be a conformal scalar, so that 
$\de_v (\pr_\mu V_\mu) = - L_v \pr_\mu V_\mu$ with $L_v$ appropriate for
a spinless field of dimension $\eta+1$, we must require
$\eta=d-1$. For a second rank tensor field 
$F_{\mu\nu}$ then similarly to (2.13) we may find
$$ 
\pr_\nu \de_v F_{\mu\nu} = - L_v (\pr_\nu  F_{\mu\nu}) + 2 (\eta -d + 1 )
b_\nu F_{\mu\nu} - 2 b_\nu F_{\nu \mu} + 2b_\mu F_{\nu\nu} \, .
\eqno (2.14) $$
for $L_v$ here acting on a vector field of dimension $\eta+1$.
The inhomogeneous terms in (2.14) disappear
if $F_{\mu\nu}=F_{[\mu\nu]}, \, {\eta=d-2}$, which justifies (1.2), or if
$F_{\mu\nu}\to T_{\mu\nu}=T_{\nu\mu}, \, T_{\mu\mu}=0$ and $\eta=d$.
If $C_{\mu\si\rho\nu}$ satisfies the conditions in (1.4) then similarly
$$ \eqalign {
\pr_\rho \de_v C_{\mu\si\rho\nu} = {}& - L_v ( \pr_\rho C_{\mu\si\rho\nu} )
+ 2 (\eta -d+1) b_\rho  C_{\mu\si\rho\nu} \, , \cr
\pr_\si \pr_\rho \de_v C_{\mu\si\rho\nu} = {}& - L_v ( \pr_\si \pr_\rho
C_{\mu\si\rho\nu} ) + 2(\eta -d+2) ( b_\rho \pr_\si C_{\mu\si\rho\nu}
+ b_\si \pr_\rho  C_{\mu\si\rho\nu} ) \, . \cr}
\eqno (2.15) $$
Hence this verifies that $\pr_\si \pr_\rho C_{\mu\si\rho\nu}$ is a traceless
tensor field if $C_{\mu\si\rho\nu}$ has dimension $d-2$ as well as 
obeying the Weyl tensor symmetries in (1.4).

Using the results in (2.6) it is straightforward to construct conformally
covariant expressions for the two point functions of quasi-primary operators.
For the field $\O$ transforming as in (2.10) and its associated conjugate
field $\bO$, which is assumed to transform as $\bO_i(x) \to  \Omega(x)^{\eta}
\bO_j (x)  D^{-1\, j} {}_{\! i}(\R (x) )$, then,
if the representation of $O(d)$ to which $\O , \bO$ belong is
irreducible, we may write in general
$$ \l \O^i (x) \, \bO_j (y) \r = {C_\O \over (s^2)^\eta} \,
D^i {}_{\! j}( I (s) ) \, , \quad s=x-y \, ,
\eqno (2.16) $$
for $C_\O$ an overall constant scale factor. Conformal invariance requires
that the two point function is zero unless both fields have the same scale
dimension. Applying this result to the conserved vector current $V_\mu$
we have
$$ 
\l V_{\mu} (x) \, V_\nu (y) \r = {C_V\over s^{2(d-1)}} \, I_{\mu \nu}
(s) \, ,
\eqno (2.17) $$
while for the energy momentum tensor $T_{\mu\nu}$
$$
\l T_{\mu \nu}(x) \, T_{\si \rho} (y)\r = {C_T\over s^{2d}} \,
\I^T{}_{\!\! \mu \nu,\si \rho} (s) \, , \quad
\I^T{}_{\!\!\mu \nu,\si \rho} (s) =  I_{\mu\alpha}(s)I_{\nu\beta}(s)
\E^T{}_{\!\! \alpha\beta,\si\rho} \, ,
\eqno (2.18) $$
where
$$
\E^T{}_{\!\!\mu\nu,\si\rho} = \half \bigl (\de_{\mu \si} \de_{\nu \rho} 
+ \de_{\mu \rho}
\de_{\nu \si} \bigl){} - {1\over d} \, \de_{\mu \nu} \de_{\si \rho}
\eqno (2.19) $$
is the projection operator onto the space of symmetric
traceless tensors so that $\I^T$ represents the corresponding inversion
tensor. Since $\pr_\mu V_\mu$ is a scalar and $\pr_\mu T_{\mu\nu}$
is a vector, if $V_\mu, T_{\mu\nu}$ have dimensions $d-1,d$, this ensures
that (2.17) and (2.18) automatically satisfy the required conservation 
equations as the two point function  of $\pr_\mu V_\mu$ and $V_\nu$ or
$\pr_\mu  T_{\mu \nu}$ and $T_{\si \rho}$ must vanish for general reasons
of conformal invariance.

The general formula for a conformally covariant three point function for
quasi-primary fields is also relatively
simple when expressed in terms of the vector $Z$\footnote{*}{This is 
entirely equivalent to the result given in \one\ in (2.13) for a convenient
choice of the arbitrary parameter $q$ there.}
$$ \eqalign {
\l \O_1^{i}(x)& \, \O_2^{j} (y) \, \O_3^{k} (z) \r  \cr
& = {1\over (x-z)^{2\eta_1}\,(y-z)^{2\eta_2}} \,
D_1^{\, i} {}_{i'} (I(x-z))
D_2^{\, j} {}_{j'} (I(y-z)) \, t_{12,3}^{i'j'k} (Z) \, . \cr }
\eqno (2.20) $$
Since the parallel transport relation (2.6) extends to arbitrary
representations it is sufficient to require that $t_{12,3}^{ijk} (Z)$ 
is a homogeneous function satisfying
$$ \eqalign {
t_{12,3}^{ijk}(\lambda Z) = {}& \lambda^{\eta_3 - \eta_1 - \eta_2}\,
t_{12,3}^{ijk}(Z) \, , \cr
D_1^{\, i} {}_{i'} (R) D_2^{\, j} {}_{j'} (R)
D_3^{\, k} {}_{ k'} (R) & \, t_{12,3}^{i'j'k'} (Z) =
t_{12,3}^{ijk}(RZ) \ \hbox {for all}\ R \in O(d) \, . \cr}
\eqno (2.21) $$
Using this with (2.9) and $I_{\si\alpha}(y-z)I_{\alpha\mu}(z-x)
= I_{\si\alpha}(y-x) I_{\alpha\mu}(X)$ the equivalent result
$$ \eqalign {
\l \O_1^{i}(x)& \, \O_2^{j} (y) \, \O_3^{k} (z) \r  \cr
& = {1\over (x-y)^{2\eta_2}\,(x-z)^{2\eta_3}} \,
D_2^{\, j} {}_{j'} (I(y-x))
D_3^{\, k} {}_{k'} (I(z-x)) \, t_{23,1}^{j'k'i} (X) \, , \cr }
\eqno (2.22) $$
is obtained where
$$
t_{23,1}^{jk\, i} (X) = (X^2)^{\eta_1 - \eta_3} D_2^{\, j} {}_{j'} (I(X))\,
t_{12,3}^{ij'k}(- X)\, .
\eqno (2.23) $$
The form of the three point function is determined then by solving (2.21).
For all fields identical then symmetry of $\l\O^{i}(x)\,\O^{j}(y)\,\O^{k}(z)\r$,
for bosonic fields, requires
$$
t^{ijk}(Z) = t^{jik}(-Z) = D^i {}_{i'} (I(Z))\, t^{ki'j}(- Z) \, ,
\eqno (2.24) $$ 
A particular solution of this condition may be found if the representation
to which $\O$ belongs allows for a completely symmetric invariant tensor
$d^{ijk}$. Using the invariance condition 
$D^i {}_{i'} (R) D^j {}_{j'} (R) D^k {}_{k'}(R)  d^{i'j'k'} = d^{ijk}$
for $R\to I(Z)$ then since $I^2=1$ (2.24) is satisfied if we take in the
expression (2.20) for this case
$$
t^{ijk}(Z) = D^k {}_{k'} (I(Z)) d^{ijk'} \, {1\over (Z^2)^{\hh \eta}} \, .
\eqno (2.25) $$

The function $t^{ijk}_{12,3}(Z)$ has a direct significance since it represents
directly the leading term in the operator product expansion. From (2.22,23)
and (2.16) it is easy to see that the leading contribution of the 
operator ${\bar \O}_3$ to the operator product of $\O_1(x) \O_2(y)$ as
$x\to y$ is given by
$$
\O^i_1(x)\O^j_2(y) \sim {1\over C_{\O_3}}t^{ijk}_{12,3}(x-y) \, 
{\bar \O}_{3\, k} (y) \, .
\eqno (2.26) $$
A generalisation \one\ of an argument due to Cardy \cardy\ shows how the 
complete three point function, as given by (2.20), may be recovered just from
the leading singular term in the operator product expansion by a simple
application of suitable conformal transformations.

In this paper the main subject of interest is the energy momentum tensor
$T_{\mu\nu}$ satisfying (1.1). This satisfies Ward identities reflecting
its role as a generator of conformal transformations \refs{\one,\pet}. The
crucial Ward identity for a correlation function for quasi-primary fields
at $y_1,y_2, \dots$ may be expressed in the form
$$
\int_S \!\! \d S_\mu \, v_\nu (x) \l T_{\mu\nu} (x) \, \O^i(y_1) \dots \r
= \l \de_v  \O^i(y_1) \dots \r \, ,
\eqno (2.27) $$
where $S$ is a surface enclosing the point $y_1$ (if 
$S$ encloses other points $y_r$ then the r.h.s is a sum of
terms involving the conformal variation of the field at each $y_r$ in turn). 
For $v$ satisfying (2.2) the l.h.s.
is invariant under smooth changes in $S$ so long as it does not cross any
of the points $y_r$. If $S$ is restricted to
a sphere surrounding the point $y_1$ with radius tending to zero we may use the
operator product expansion in the form \cardy\
$$ T_{\mu\nu}(x)\, \O(y) \sim A_{\mu\nu} (r) \O(y) + 
B_{\mu\nu\lambda} (r)\pr_\lambda \O(y) \, , \quad r=x-y \, ,
\eqno (2.28) $$
and then (2.27) in conjunction with (2.11) for $\de_v \O$, since now 
$\d S_\mu = |r|^{d-1}\hr_\mu \d\Omega_{\hr}$ for 
$\hr_\mu = r_\mu/|r|, \, {|r|=\sqrt{ r^2}}$, requires the conditions
$$ \eqalign {
\int \!\! \d\Omega_{\hr} \, \hr_\mu A_{\mu\nu} (\hr) = {}& 0 \, , \qquad 
\int \!\! \d\Omega_{\hr} \, \hr_\mu B_{\mu\nu\lambda} (\hr) =
- \de_{\nu\lambda} \, , \cr
\int \!\! \d\Omega_{\hr} \, \hr_\mu \hr_\omega A_{\mu\nu} (\hr) = {}&
-{\eta\over d}\, \de_{\omega\nu} + \half s_{\omega\nu} -
{\hat C}_{\omega\nu} \, , \quad {\hat C}_{\omega\nu} = {\hat C}_{\nu\omega} \, ,
\quad {\hat C}_{\nu\nu} = 0 \, . \cr}
\eqno (2.29) $$
$s_{\omega\nu}$ is the usual generator of $O(d)$
in the representation defined by $\O$ while ${\hat C}_{\omega\nu}$ is
an $O(d)$ invariant (satisfying $[s_{\mu\nu} , {\hat C}_{\omega\lambda}]
+ (s_{\mu\nu})_{\omega\lambda,\omega'\lambda'}{\hat C}_{\omega'\lambda'} = 0$)
and is otherwise undetermined by the Ward identity.

As mentioned above, the operator product
coefficients are related to the appropriate three point function. We may
therefore find the three point function of the energy momentum tensor, a
general operator $\O$ and its conjugate $\bO$ by applying (2.20),
$$ \eqalign {
\l T_{\mu\nu}(x)& \, \O^{i} (y) \, \bO_j (z) \r  \cr
& = C_\O \, {1\over (x-z)^{2d}\,(y-z)^{2\eta}} \,
\I^T{}_{\!\! \mu \nu,\si \rho} (x-z)
D^i {}_{i'} (I(y-z)) \, A_{\si\rho}{}^{\! i'}{}_{\! j} (Z) \, . \cr }
\eqno (2.30) $$
since the leading term in the operator product expansion (2.28),
according to (2.26), determines the appropriate form for the
relevant function $t^{ijk}_{12,3}(Z)$.
In accord with (2.21) $A_{\mu\nu} (Z) = \rO(Z^{-d})$ and also satisfies
the appropriate corresponding rotational covariance condition.
The conservation equation (1.1) also leads directly to
$\pr_\mu A_{\mu\nu} (Z)=0$ for $Z\ne 0$. As a consequence of
the three point function being fully determined by the leading operator
product expansion coefficient $A_{\mu\nu}$ it is possible by careful
expansion to evaluate the next leading term $B_{\mu\nu\lambda}$ in (2.28)
in terms of $A_{\mu\nu}$. Using this result, given in \one, it is easy to
verify that the condition on $B_{\mu\nu\lambda}$ in (2.29) holds so long as
the relations for $A_{\mu\nu}$ are satisfied. If $A_{\mu\nu}(x)$ is
regarded as a distribution on $\bR^d$ the Ward identity results (2.29) may be
rewritten equivalently in the form
$$ \eqalign {
\pr_\mu A_{\mu\nu}(x) = {}& \Big ( {\eta\over d}\, \de_{\nu\lambda} 
+ C_{\nu\lambda} + \half s_{\nu\lambda}\Big ) \pr_\lambda \de^d(x) \, , \cr
A_{\mu\mu}(x) = {}& C_{\mu\mu} \de^d(x) \, , \qquad
\pr_\mu B_{\mu\nu\lambda} (x) = -\de_{\nu\lambda} \de^d(x) \, , \cr}
\eqno (2.31) $$
where $C_{\mu\nu}{}^i{}_{\!j}=C_{\nu\mu}{}^i{}_{\!j}$ is a general $O(d)$
invariant. This term in the
result for $\pr_\mu A_{\mu\nu}$ and $A_{\mu\mu}$ reflects the arbitrariness
due to regularisation dependent ambiguities in $A_{\mu\nu}(x)$ proportional to
$\de^d(x)$ when it is extended to a well defined distribution\footnote{*}{If
we define
$\pr_\mu A_{\mu\nu}(x) \equiv \lim_{\omega \to 0} \pr_\mu \big ( x^{2\omega}
A_{\mu\nu}(x) \big )$ then in (2.31) $C_{\mu\nu} = {\hat C}_{\mu\nu}$}.
The ambiguity represented by $C_{\mu\nu}$ arises essentially from the 
freedom of definition of the three point function (2.30) up to changes of
the form
$$ \eqalign{
\l T_{\mu\nu}(x) \, \O (y) \, \bO (z) \r \to {}&
\l T_{\mu\nu}(x) \, \O (y) \, \bO (z) \r \cr
& + \de^d(x-y) C_{\mu\nu}
\l \O (y) \, \bO (z) \r + \de^d(x-z) \l \O (y) \, \bO (z) \r C_{\mu\nu} \, .\cr}
\eqno (2.32) $$
Corresponding to (2.31) we may then find a general expression for the Ward
identities for the three point function in (2.30)
$$ \eqalign{ \!\!\!\! \!\!\!\!\!\!\!\!
\pr^x{}_{\! \mu} \l T_{\mu\nu}(x) \, \O (y) \, \bO (z) \r = {}&
\pr_\nu \de^d(x-y) \, \l \O(x) \, \bO (z) \r + \pr_\nu \de^d(x-z)
\, \l \O(y) \, \bO(x) \r \cr
& + \pr^x{}_{\! \lambda} \Big \{ \de^d(x-y) \Big ( {\eta-d\over d}
\de_{\nu\lambda} + C_{\nu\lambda} + \half s_{\nu\lambda}\Big )
\l \O(x) \, \bO (z) \r \cr
& \quad \quad \ +  \de^d(x-z)\l \O(y) \, \bO(x) \r 
\Big ( {\eta-d\over d} \de_{\nu\lambda} + C_{\nu\lambda}
- \half s_{\nu\lambda}\Big ) \Big \} \, , \cr
\l T_{\mu\mu}(x) \, \O (y) \, \bO (z) \r  = {}& 
\de^d(x-y) C_{\mu\mu}
\l \O (y) \, \bO (z) \r + \de^d(x-z) \l \O (y) \, \bO (z) \r C_{\mu\mu} \, .\cr}
\eqno (2.33) $$
For suitable choices of $C_{\mu\nu}$ these identities are identical with
standard Ward identities whose derivations may depend on particular 
regularisation schemes.

As an illustration we discuss the Ward identities for the three point
function of the energy momentum tensor itself which may be derived by
functional differentiating (1.6) and then restricting to flat space,
$$ \eqalignno {
\pr^x{}_{\! \mu} \l T_{\mu\nu} (x) T_{\si\rho} (y) 
T_{\alpha \beta} (z) \r &
=  \pr_\nu \de^d(x-y) \, \l T_{\si\rho} (x) T_{\alpha \beta} (z)\r \cr
& +\pr_\si \bigl ( \de^d(x-y) \l T_{\nu\rho} (x) T_{\alpha \beta} (z)\r \bigl)
{} + \pr_\rho \bigl ( \de^d(x-y) \l T_{\si\nu} (x) T_{\alpha \beta} (z)\r 
\bigl ) \cr 
&{}\quad + y, \, \si, \, \rho \leftrightarrow z, \, \alpha,  \, \beta \, . 
& (2.34) \cr}
$$
This form for the Ward identity is in exact agreement with (2.33) if we take
$(C_{\nu\lambda})_{\si\rho,\si'\rho'} = \E^T{}_{\!\!\si\rho,\nu\ep}
\E^T{}_{\!\!\lambda\ep,\si'\rho'} + \nu \leftrightarrow \lambda$. Since
$(C_{\mu\mu})_{\si\rho,\si'\rho'} = 2 \E^T{}_{\!\!\si\rho,\si'\rho'}$ the
corresponding trace identity becomes
$$
\l T_{\mu\mu} (x)  \, T_{\si\rho} (y) \, T_{\alpha \beta} (z) \r
= 2  \bigl (\de^d(x-y)+\de^d(x-z)\bigl)\,
\l T_{\si\rho} (y) \, T_{\alpha \beta} (z) \r \, ,
\eqno (2.35) $$
which is identical with (1.9) for $d=4$
if the anomaly terms proportional to $\beta_{a,b}$ are dropped.

The presence of the trace anomaly leads to extra terms in the equations
such as (2.27) expressing conformal invariance when $d=4$. For instance 
if we integrate
$\pr^x{}_{\! \mu}\big ( v_\nu(x)  \l T_{\mu\nu} (x) T_{\si\rho} (y)
T_{\alpha \beta} (z) \r \big )$ over all $x$, dropping the surface term for
$|x|\to \infty$, then, if $v_\mu(x)$ is assumed to satisfy the conditions
for an infinitesimal conformal transformation as in (2.2), using (2.34)
for $d=4$ and (1.9) gives
$$ 
\l L_v T_{\si\rho} (y) \, T_{\alpha \beta} (z) \r
+ \l T_{\si\rho} (y)\, L_v  T_{\alpha \beta} (z) \r
= - 32 \beta_a\,
\E^C{}_{\!\! \si\ep\eta\rho,\alpha\gamma\de\beta}\, \pr_{\ep} \pr_{\eta}
\si_v(y) \pr_{\gamma} \pr_{\de} \de^4(y-z) \, ,
\eqno (2.36) $$
where $L_v$ is defined as in (2.11) with the scale dimension $\eta=d$. There
is no term involving $\beta_b$ in this case since $\pr\pr \si_v=0$. This
result shows that $\beta_a$ must be related to the coefficient $C_T$ of
the energy momentum tensor two point function as exhibited in (2.18).
To obtain the exact relationship requires a careful treatment of the short
distance singularities. From (2.19) we may write
$$ \eqalign { \!\!\!\!\!
\l T_{\si\rho} (y) \, T_{\alpha \beta} (z) \r ={}& C_T \,
{\I^T{}_{\!\! \si \rho,\alpha\beta}(y-z)\over (y-z)^{2d}} \cr
= {}& {C_T \over (d-3)(d-2)d(d+1)}
\E^C{}_{\!\! \si\ep\eta\rho,\alpha\gamma\de\beta}\, \pr^y{}_{\!\!\ep} 
\pr^y{}_{\!\!\eta}
\pr^z{}_{\!\!\gamma} \pr^z{}_{\!\!\de} {1\over (y-z)^{2(d-2)}} \, , \cr}
\eqno (2.37) $$
where the second form ensures that the conservation equations are trivially
satisfied although it is no longer manifestly conformally covariant. When
$d=4$ the singularity as $y\to z$ is not integrable so some regularisation,
denoted by $\R$, is necessary
$$
\l T_{\si\rho} (y) \, T_{\alpha \beta} (z) \r = {C_T\over 40}\,
\E^C{}_{\!\! \si\ep\eta\rho,\alpha\gamma\de\beta}\, \pr^y{}_{\!\!\ep} 
\pr^y{}_{\!\!\eta}
\pr^z{}_{\!\!\gamma} \pr^z{}_{\!\!\de}\, \R {1\over (y-z)^4} \, .
\eqno (2.38) $$
A convenient definition in this case is to take
$$
\R {1\over x^4} = \Big ( {\mu^{2\omega} \over x^{2(2-\omega)}} - {\pi^2
\over \omega} \de^4(x) \Big ) \Big |_{\omega \to 0} = 
-\pr^2 \, {1\over 4x^2}\big ( \ln \mu^2x^2 + 1 \big )\, ,
\eqno (2.39) $$
with $\mu$ an arbitrary regularisation scale and where $1/x^{2(2-\omega)}$
is defined as an analytic function of $\omega$, the limit $\omega\to 0$
may be taken as a distribution after subtracting the pole. In the second
expression for $\R(1/x^4)$ in (2.39) the limit is found explicitly following
the method of differential regularisation \diff. 
Using the representation (2.37) or (2.38) the conformal
variation of the energy momentum tensor two point function
may be determined with the aid of the essential result obtained in (2.15),
$L_v(\pr_\ep \pr_\eta C_{\si\ep\eta\rho}) 
= \pr_\ep \pr_\eta(L_vC_{\si\ep\eta\rho})$ if $C_{\si\ep\eta\rho}$ 
has dimension $d-2$ and obeys the symmetry conditions of (1.4).
The conformal variation given by $L_v$ applied to (2.38) then reduces to
a term involving
$$
\big ( v(y){\cdot \pr^y} +  v(z){\cdot \pr^z} + 2\si_v(y) + 2\si_v(z) \big )
\R {1\over (y-z)^4} = 2\pi^2 \si_v(y) \de^4(y-z) \, ,
\eqno (2.40) $$
and in addition terms representing the associated conformal variation of
$\E^C{}_{\!\! \si\ep\eta\rho,\alpha\gamma\de\beta}$ which is given by the
action of the rotation generator corresponding to $\pr_{[\mu}v_{\nu]}(y)$ for
the indices $\si\ep\eta\rho$ and to $\pr_{[\mu}v_{\nu]}(z)$ acting on the
indices $\alpha\gamma\delta\beta$.
Using that $\E^C$ is an invariant tensor we then obtain
$$ \eqalignno { \!\!\!\!
\l L_v T_{\si\rho} (y) \, T_{\alpha \beta} (& z) \r
+ \l T_{\si\rho} (y)\, L_v  T_{\alpha \beta} (z) \r 
= {C_T \pi^2\over 20}\E^C{}_{\!\! \si\ep\eta\rho,\alpha\gamma\de\beta}\, 
\pr^y{}_{\!\!\ep} \pr^y{}_{\!\!\eta} \pr^z{}_{\!\!\gamma} \pr^z{}_{\!\!\de} 
\big ( \si_v(y) \de^4(y-z) \big ) \cr 
& +  {3C_T \over 40}b_\lambda \, 
\pr^y{}_{\!\!\ep} \pr^y{}_{\!\!\eta} \pr^z{}_{\!\!\gamma} \pr^z{}_{\!\!\de}
\pr^y{}_{\!\![\lambda}\big (\E^C{}_{\!\! \si\ep]\eta\rho,\alpha\gamma\de\beta}
+ \E^C{}_{\!\! \rho\eta]\ep\si,\alpha\gamma\de\beta} \big )
{1\over (y-z)^2} \, . & (2.41) \cr}
$$
The symmetries of $\E^C$ ensure that the second line, depending on $b_\lambda$,
vanishes and comparing with (2.36) gives finally
$$
\beta_a = - {\pi^2\over 640}\, C_T \, .
\eqno (2.42) $$

In a similar fashion the vector current two point function in (2.17)
may be written for $d=4$ in the form
$$
\l V_{\mu} (x) \, V_\nu (y) \r = - {C_V \over 6}\,\E^F{}_{\!\!\mu\si,\nu\rho}\,
\pr^x{}_{\!\!\si} \pr^y{}_{\!\!\rho} \, \R {1\over (x-y)^4} \,, \quad 
\E^F{}_{\!\!\mu\si,\nu\rho} = \half \big ( \de_{\mu\nu}\de_{\si\rho}
- \de_{\mu\rho} \de_{\si\nu}\big ) \, .
\eqno (2.43) $$
Following a similar analysis to the above we may find
$$ \eqalign { \!\!\!\!
\l L_v  V_{\mu} (x) \, V_\nu (y) \r + \l V_{\mu} (x) \, L_v V_\nu (y) \r ={}&
- {C_V \pi^2\over 3} \,\E^F{}_{\!\!\mu\si,\nu\rho} \pr^x{}_{\!\!\si}
\pr^y{}_{\!\!\rho} \big ( \si_v(x) \de^4 (x-y) \big ) \cr
& - {C_V\over 2} b_\lambda \pr^x{}_{\!\!\si} \pr^y{}_{\!\!\rho}
\pr^x{}_{\!\![\lambda} \E^F{}_{\!\!\mu\si],\nu\rho} {1\over (x-y)^2} \, .\cr}
\eqno (2.44) $$
The second term vanishes identically as previously and the result may be
compared with the consequences of the trace anomaly in (1.7) to give
$$
\kappa = {\pi^2\over 6} C_V \, .
\eqno (2.45) $$
In consequence of (2.42,45) the trace anomaly coefficients $\kappa$ and
$\beta_a$ determine the scale of the vector current and energy momentum tensor
two point functions in the conformal limit.
\bigskip
\leftline{\bigbf 3 Axial Anomaly}
\medskip
As a first application of these results\footnote{*}{A discussion of the
axial anomaly in configuration space is also contained in ref. \free.}
 we consider the archetypal
situation in which an anomaly is present, the three point function of
two vector currents $V$ and an axial current $A$. In this section we set
$d=4$ throughout since the $\vep$ tensor plays an essential role. The conserved
vector and axial currents then have dimension 3. We impose  manifest
conservation of the vector currents by requiring that the three point function
is written in the form
$$
\l V_\mu (x) V_\nu(y) A_\omega (z) \r = \pr^x_{\, \si} \pr^y_{\, \rho}
\Gamma^{FFA}_{\mu\si,\nu\rho ,\omega}(x,y,z) \, , \quad
\Gamma^{FFA}_{\mu\si,\nu\rho ,\omega} = \Gamma^{FFA}_{[\mu\si],\nu\rho, \omega}
= \Gamma^{FFA}_{\mu\si,[\nu\rho] ,\omega} \, ,
\eqno (3.1) $$
with the symmetry condition $\Gamma^{FFA}_{\mu\si,\nu\rho, \omega}(x,y,z)
= \Gamma^{FFA}_{\nu\rho,\mu\si,\omega}(y,x,z)$.  From the 
conservation equation for the axial current we must also require 
that $\pr^z_{\, \omega} \Gamma^{FFA}_{\mu\si,\nu\rho ,\omega}(x,y,z) = 0 $ 
for non coincident points $x,y,z$. Assuming conformal invariance, with
$\Gamma^{FFA}_{\mu\si,\nu\rho,\omega}(x,y,z)$ the three point function for
an antisymmetric tensor field of
dimension 2 at $x,y$, and also the required parity properties leads to
an essentially unique solution,
$$
\Gamma^{FFA}_{\mu\si,\nu\rho,\omega}(x,y,z) = C \, {\I^F{}_{\!\!\mu\si,\mu'\si'}
(x-y)\over \big ((x-y)^2\big )^2} \, \vep_{\mu'\si'\nu\rho} \,Z^2 Z_\omega \, ,
\eqno (3.2) $$
for $\I^F$ representing inversions on antisymmetric tensors and defined by
$$
\I^F{}_{\!\!\mu\si,\nu\rho} = I_{\mu\mu'} I_{\si\si'}
\E^F{}_{\!\!\mu'\si',\nu\rho}\, ,
\eqno (3.3) $$
with $\E^F$ defined in (2.43).
The symmetry condition for $x\leftrightarrow y$, when $Z\to -Z$, follows
from $\det I = -1$ which implies $\I^F{}_{\!\!\mu\si,\mu'\si'}
\vep_{\mu'\si'\nu\rho}= - \I^F{}_{\!\!\nu\rho,\nu'\rho'}\vep_{\mu\si\nu'\rho'}$.
By calculating the derivatives we may find an expression equivalent to that
in refs.\refs{\schreier,\crewther}
$$
\l V_\mu (x) V_\nu(y) A_\omega (z) \r = 4C \, { I_{\mu\mu'}(x-z)
I_{\nu\nu'}(y-z) \over \big( (x-z)^2 (y-z)^2 \big )^3} \, \vep_{\mu'\nu'
\lambda\omega} {Z_\lambda \over (Z^2)^2} \, ,
\eqno (3.4) $$
which is of the standard form given by (2.20). 

The expression given by
(3.4) is non integrable as a function on $\bR^8$, with coordinates 
$x-z,y-z$, due to the singular behaviour as $x-z\sim y-z \to 0$.
The Fourier transform is therefore ill defined without regularisation (this
corresponds to the linear divergence of the usual triangle graph). However
$\Gamma^{FFA}_{\mu\si,\nu\rho,\omega}(x,y,z)$ as given in (3.2) has no non
integrable short distance singularities and represents a well defined
distribution on $\bR^8$ (the apparent sub-divergence for $x-y\to 0$
in (3.2) is absent since $Z\propto x-y$ in this limit). 
Thus the divergence of the axial current in (3.1) may be unambiguously
calculated. To achieve this we first use from (2.7)
$$  
\pr^z{}_{\! \omega} (Z^2 Z_\omega) = 2\pi^2 \big ( \de^4(y-z) - 
\de^4(x-z) \big ) \, .
\eqno (3.5) $$
The remaining derivatives can be evaluated with the aid of
$$ 
\pr_\si\, {\I^F{}_{\!\!\mu\si,\nu\rho}(x)\over x^{2\lambda}} = 
{2-\lambda\over 2\lambda} \big ( \de_{\si\rho}\de_{\mu\nu} -
\de_{\si\nu}\de_{\rho\mu}\big ) \pr_\si {1\over x^{2\lambda}}
\to \half  {\pi^2} \E^F{}_{\!\!\mu\si,\nu\rho}
\pr_\si \de^4(x) \ \ \hbox{as} \ \ 
\lambda \to 2 \, ,
\eqno (3.6) $$
which depends on the limit, as in (2.39),
$$
{1\over (x^2)^{2-\omega}} \sim {\pi^2 \over \omega}\, \de^4(x) \ \ \ 
\hbox{for} \ \ \ \omega\to 0 \, .
\eqno (3.7) $$
Using the result (3.6) for the $\pr^x{}_{\! \si}$ or $\pr^y{}_{\!\rho}$
derivatives in association with the first or second terms on the r.h.s. of
(3.5) we thereby obtain
$$
\pr^z{}_{\!\! \omega} \l V_\mu (x) V_\nu(y) A_\omega (z) \r =
2\pi^4 C \, \pr^x{}_{\!\! \si} \pr^y{}_{\!\! \rho} \big ( \vep_{\mu\si\nu\rho}
\de^4(x-z)\de^4(y-z)\big ) \, ,
\eqno (3.8) $$
which is the standard expression for the anomalous divergence in this case.

For free fermions it is easy to determine the overall coefficient using
$$
V_\mu = {\bar \psi}\gamma_\mu \psi \, , \quad A_\mu = 
{\bar \psi}\gamma_\mu \gamma_5 \psi \, , \quad
\l \psi(x) {\bar \psi}(0) \r = {\gamma{\cdot x} \over 2\pi^2 (x^2)^2} \, ,
\eqno (3.9) $$
with also ${\rm tr} (\gamma_\alpha \gamma_\beta \gamma_\gamma\gamma_\delta
\gamma_5 ) = 4 \vep_{\alpha\beta\gamma\delta}$. The result for the usual
triangle graphs coincides with (3.4) if
$$
C={1\over 4\pi^6} \, ,
\eqno (3.10) $$
although the calculation may be simplified by restricting $x,y,z$ to
be collinear such as along the $1$-direction. The formula (3.8) for the
anomaly coefficient then corresponds with that expected from (1.5).

Of course far more sophisticated methods for determining the anomaly in
$\pr_\omega \l A_\omega \r$, on general backgrounds, are known but the above
derivation perhaps makes natural the essential 
independence of the result of any 
regularisation procedure once vector current conservation has been imposed. 
\bigskip
\leftline{\bigbf 4 Conserved Currents and Energy Momentum Tensor}
\medskip
An example which is less involved than dealing with the energy momentum 
alone concerns the three point function of two vector currents and the energy 
momentum tensor. Previously it was shown \one\ that there are two possible
linearly independent forms assuming conformal invariance. Since conservation
of the vector currents does not lead to any non trivial Ward identities
we here restrict attention to the form
$$  
\l V_\mu (x) V_\nu(y) T_{\alpha\beta} (z) \r = \pr^x_{\, \si} \pr^y_{\, \rho}
\Gamma^{FFT}_{\mu\si,\nu\rho,\alpha\beta}(x,y,z) \, , 
\eqno (4.1) $$
with $\Gamma^{FFT}_{\mu\si,\nu\rho,\alpha\beta}$ antisymmetric on $\mu\si$
and $\nu\rho$ but symmetric and traceless on $\alpha\beta$.
Applying conformal invariance we may assume in accord with (2.16)
$$
\Gamma^{FFT}_{\mu\si,\nu\rho,\alpha\beta}(x,y,z) =
{\I^F{}_{\!\!\mu\si,\mu'\si'}(x-z)\I^F{}_{\!\!\nu\rho,\nu'\rho'}(y-z)
\over \big ((x-z)^2(y-z)^2\big )^{d-2}}\, t^{FFT}_{\,\mu'\si', \nu'\rho',
\alpha\beta} (Z) \, ,
\eqno (4.2) $$
where $t^{FFT}$ can be expanded in general in the form
$$ \eqalign {
t^{FFT}_{\,\mu\si,\nu\rho,\alpha\beta} (Z) = {}&
A \, \E^F{}_{\!\! \mu\si,\lambda\ep} \E^F{}_{\!\! \nu\rho,\lambda\eta}
\E^T{}_{\!\! \ep \eta,\alpha\beta} {1\over (Z^2)^{\hh d - 2}} \cr
& + B\, \E^F{}_{\!\! \mu\si,\kappa\ep} \E^F{}_{\!\! \nu\rho,\lambda\eta}
\E^T{}_{\!\!\kappa\lambda,\alpha\beta} {Z_\ep Z_\eta\over (Z^2)^{\hh d - 1}} \cr
& + C\, \E^F{}_{\!\! \mu\si,\lambda\ep} \E^F{}_{\!\! \nu\rho,\lambda\eta}
\big ( \E^T{}_{\!\! \ep \kappa,\alpha\beta} Z_\eta +
\E^T{}_{\!\! \eta \kappa,\alpha\beta} Z_\ep \big ) 
{Z_\kappa\over (Z^2)^{\hh d - 1}} \cr
& + D\, \E^F{}_{\!\! \mu\si,\nu\rho} \Big ( {Z_\alpha Z_\beta \over Z^2}
- {1\over d} \de_{\alpha\beta} \Big ) {1\over (Z^2)^{\hh d - 2}} \cr
& + E\, \E^F{}_{\!\! \mu\si,\lambda\ep} \E^F{}_{\!\! \nu\rho,\lambda\eta}
Z_\ep Z_\eta \Big ( {Z_\alpha Z_\beta \over Z^2}
- {1\over d} \de_{\alpha\beta} \Big ) {1\over (Z^2)^{\hh d - 1}} \, . \cr}
\eqno (4.3) $$
It is also useful to rewrite (4.2) as in (2.22) alternatively as
$$
\Gamma^{FFT}_{\mu\si,\nu\rho,\alpha\beta}(x,y,z) =
{\I^F{}_{\!\!\nu\rho,\nu'\rho'}(y-x) \I^T{}_{\!\! \alpha\beta,\alpha'\beta'}
(z-x) \over \big ( (y-x)^2\big )^{d-2}\big ( (z-x)^2\big )^{d}}\, 
{\tilde t}^{FFT}_{\,\mu\si, \nu'\rho',\alpha'\beta'} (X) \, ,
\eqno (4.4) $$
where now, applying (2.23) in this case,
$$ \eqalign { \!\!\!\!\!\!
{\tilde t}^{FFT}_{\,\mu\si,\nu\rho,\alpha\beta} (X) = {}&
{1\over (X^2)^2} \I^F{}_{\!\!\nu\rho,\nu'\rho'}(X) 
t^{FFT}_{\,\mu\si, \nu'\rho',\alpha\beta} (-X) \cr
={}& A \, \E^F{}_{\!\! \mu\si,\lambda\ep} \E^F{}_{\!\! \nu\rho,\lambda\eta}
\E^T{}_{\!\! \ep \eta,\alpha\beta} {1\over (X^2)^{\hh d }} \cr
& -(2A+B)\, \E^F{}_{\!\! \mu\si,\kappa\ep} \E^F{}_{\!\! \nu\rho,\lambda\eta}
\E^T{}_{\!\! \kappa\lambda,\alpha\beta}{X_\ep X_\eta\over (X^2)^{\hh d + 1}} \cr
& + C\, \E^F{}_{\!\! \mu\si,\lambda\ep} \E^F{}_{\!\! \nu\rho,\lambda\eta}
\E^T{}_{\!\! \kappa\eta,\alpha\beta} {X_\ep X_\kappa\over (X^2)^{\hh d + 1}} \cr
& - (2A+C)\, \E^F{}_{\!\! \mu\si,\lambda\ep} \E^F{}_{\!\! \nu\rho,\lambda\eta}
\E^T{}_{\!\! \kappa\ep,\alpha\beta} {X_\eta X_\kappa\over (X^2)^{\hh d + 1}}\cr
& + D\, \E^F{}_{\!\! \mu\si,\nu\rho} \Big ( {X_\alpha X_\beta \over X^2}
- {1\over d} \de_{\alpha\beta} \Big ) {1\over (X^2)^{\hh d }} \cr
& -(2C+4D+E)\, \E^F{}_{\!\! \mu\si,\lambda\ep} \E^F{}_{\!\! \nu\rho,\lambda\eta}
X_\ep X_\eta \Big ( {X_\alpha X_\beta \over X^2}
- {1\over d} \de_{\alpha\beta} \Big ) {1\over (X^2)^{\hh d + 1}} \, . \cr}
\eqno (4.5) $$
Imposing the conservation equation
$$
\pr^z_{\, \alpha} \Gamma^{FFT}_{\mu\si,\nu\rho,\alpha\beta}(x,y,z) = 0 
\ \ \Rightarrow \ \ \pr^X_{\, \alpha}
{\tilde t}^{FFT}_{\,\mu\si,\nu\rho,\alpha\beta} (X) = 0 \, ,
\eqno (4.6) $$
and this gives rise to the single condition
$$
K\equiv 2(2C+4D+E) + (d+2)B - (d+2)(d-4) A = 0 \, .
\eqno (4.7) $$

{}From (4.2) it is easy to determine the singular behaviour as $z\to y$
since then
$$
\Gamma^{FFT}_{\mu\si,\nu\rho,\alpha\beta}(x,y,z) \sim
{\I^F{}_{\!\!\mu\si,\mu'\si'}(x-y)\over (x-y)^{2(d-2)}} \, 
{\tilde t}^{FFT}_{\,\mu'\si', \nu\rho,\alpha\beta} (y-z) \, ,
\eqno (4.8) $$
and similarly as $z\to x$. From (4.5) this is $\rO((z-y)^{-d})$, which gives
potential logarithmic divergences on integration on $\bR^d$. To analyse 
these singularities more closely we use generalisations of (3.7) for arbitrary
$d$
$$ \eqalign {
{1\over (x^2)^{\hh d-\omega}} \sim{}&{1\over 2\omega}\, S_d \de^d(x) \, , \qquad
{x_\mu x_\nu\over (x^2)^{\hh d -\omega+1}} \sim {1\over 2\omega}\, 
{1\over d} \de_{\mu\nu} S_d \de^d(x) \, , \cr
{x_\mu x_\nu x_\si x_\rho\over (x^2)^{\hh d-\omega+2}}\sim{}&{1\over 2\omega}\, 
{1\over d(d+2)} \big ( \de_{\mu\nu}\de_{\si\rho} +
\de_{\mu\si}\de_{\nu\rho}+ \de_{\mu\rho}\de_{\nu\si}\big )  S_d \de^d(x) \, ,
\quad S_d = {2\pi^{\hh d}\over \Gamma(\half d)} \, , \cr}
\eqno (4.9) $$
where the singularity is represented by a pole in $\omega$. Applying these
results to the detailed expression (4.5) we may then find
$$
x^{2 \omega}{\tilde t}^{FFT}_{\,\mu\si, \nu\rho,\alpha\beta} (x)
\sim - {1\over 2\omega}\, {K\over d(d+2)} \, 
\E^F{}_{\!\! \mu\si,\lambda\ep} \E^F{}_{\!\! \nu\rho,\lambda\eta}
\E^T{}_{\!\! \ep \eta,\alpha\beta} \, S_d \de^d(x) \, ,
\eqno (4.10) $$
with $K$ as in (4.7). Hence imposing the condition $K=0$ is sufficient to
ensure that there are no non integrable singularities for $x,y \to z$ in
$\Gamma^{FFT}_{\mu\si,\nu\rho,\alpha\beta}(x,y,z)$ and that for general $d$
it can be extended to a well defined distribution although there are
ambiguities up to terms proportional $\de^d(x-z)$ or $\de^d(y-z)$ as in (2.33).

For non coincident points we may write from (4.1,2)
$$
\l V_\mu (x) V_\nu(y) T_{\alpha\beta} (z) \r =
- {I_{\mu\mu'}(x-z)I_{\nu\nu'}(y-z)
\over \big ((x-z)^2(y-z)^2\big )^{d-1}}\, \pr_\si^{\vphantom X}
\pr_\rho ^{\vphantom X} t^{FFT}_{\,\mu'\si, \nu'\rho,\alpha\beta} (Z) \, ,
\eqno (4.11)$$
where
$$ \eqalign {
- \pr_\si^{\vphantom {X}}\pr_\rho^{\vphantom {X}} 
t^{FFT}_{\,\mu\si, \nu\rho,\alpha\beta} (Z) = {}&
I \bigg (  \E^T{}_{\!\! \mu \nu ,\alpha\beta}
+ \half d(d-2) {Z_\mu Z_\nu \over Z^2}\Big ( {Z_\alpha Z_\beta \over Z^2}
- {1\over d} \de_{\alpha\beta} \Big ) \bigg ) {1\over (Z^2)^{\hh d -1}} \cr
& + J \bigg ( Z_\mu  \E^T{}_{\!\! \nu \lambda ,\alpha\beta} +
Z_\nu  \E^T{}_{\!\! \mu \lambda ,\alpha\beta} \big ) 
{Z_\lambda \over (Z^2)^{\hh d}} \cr
& - \big ( J - \half (d-2) I \big ) \, \de_{\mu\nu}
\Big ( {Z_\alpha Z_\beta \over Z^2}
- {1\over d} \de_{\alpha\beta} \Big ){1\over (Z^2)^{\hh d -1}} \, ,\cr
I = C+D - \half B \, , \quad & J = \quar \big ( 2E + (d-4)(d-2) A 
- (d-2)(B + 2C + 4D ) \big ) \, . \cr}
\eqno (4.12) $$
Thus it is clear that there are just two linearly independent forms for this
conformal invariant three point function.

When $d=4$ the resulting expression given by (4.2,3) is also ill defined
as a distribution on $\bR^8$ due to the short distance singularity when $x,y,z$ 
are all coincident becoming non integrable.
This divergence is manifested, for general $d$, by a pole in
$\vep = 4-d$. For $\Gamma^{FFT}_{\mu\si,\nu\rho,\alpha\beta}(x,y,z) $ the
pole involves just delta functions without derivatives and the tensorial
structure of the residue has an essentially unique form dictated by symmetry
requirements and assuming that it is traceless on $\alpha\beta$. 
Hence we may write in general
$$
\Gamma^{FFT}_{\mu\si,\nu\rho,\alpha\beta}(x,y,z) \sim {R\over \vep} \,
\E^F{}_{\!\! \mu\si,\lambda\ep} \E^F{}_{\!\! \nu\rho,\lambda\eta}
\E^T{}_{\!\! \ep \eta,\alpha\beta} \, S_d{}^{\! 2} \de^d(x-z) \de^d(y-z) \, .
\eqno (4.13) $$
To determine the coefficient $R$ we make use of the formula \curve\
$$
{1\over (x-y)^{2\lambda_3} (y-z)^{2\lambda_1} (z-x)^{2\lambda_2}} \sim
{1\over d - \sum_\alpha \lambda_\alpha} \, \prod_\alpha {\Gamma(\half d -
\lambda_\alpha ) \over \Gamma ( \lambda_\alpha ) } \, 
{\pi^d \over \Gamma (\half d)} \de^d(x-z) \de^d(y-z) \, ,
\eqno (4.14) $$
which exhibits the the singularity at coincident points as a singularity
in $\sum_\alpha \lambda_\alpha$. The poles which are present when
$\lambda_\alpha - \half d =0,1,\dots$ correspond to
sub-divergences for these $\lambda_\alpha$ for two points becoming
coincident, as expected according to (4.9). As given by (4.2,3)
$\Gamma^{FFT}_{\mu\si,\nu\rho,\alpha\beta}(x,y,z)$ has no subdivergences
for $x\to y$ or, subject to (4.7), if $x,y\to z$.
To apply the result (4.14) directly
to determine $R$ in (4.13) it is necessary to contract
indices to avoid complicated tensorial expressions. Hence instead of (4.13)
it is sufficient to analyse just
$$
\Gamma^{FFT}_{\mu\si,\mu\rho,\si\rho}(x,y,z) \sim {R\over \vep} \,
{\ts {1\over 8}}(d-1)(d-2)(d+2) \, S_d{}^{\! 2} \de^d(x-z) \de^d(y-z) \, .
\eqno (4.15) $$
Using (4.2,3) the left hand side of (4.15) may be expressed as a sum of terms
of the form exhibited in (4.14) with $\sum_\alpha \lambda_\alpha = {3\over 2}
d-2$ where in individual terms $\lambda_\alpha = \half d + n_\alpha$ for
$n_\alpha$ an integer.
When $n_\alpha=0,1,\dots$ for some $\alpha$ the
formula (4.14) is singular. In order to avoid these problems
we introduce an additional factor
$(x-y)^{2\omega_3}(x-z)^{2 \omega_2} (y-z)^{2 \omega_1}$ so that
$\lambda_\alpha \to \lambda_\alpha - \omega_\alpha$ and the singularities
now correspond to poles in $\omega_\alpha$. The poles in $\omega_3$
cancel when terms corresponding to each coefficient $A,B,C,D,E$ in (4.3)
are combined, so long as the residues given
by (4.14) are evaluated at the pole $\half d = 2 + \sum_\alpha \omega_\alpha$,
since each term in (4.2,3) has no non integrable singularity for $x\to y$
for $d\approx 4$.
After taking the limit $\omega_3\to 0$ the poles in $\omega_{1,2}$ also
cancel if the condition (4.7), which eliminates
the sub-divergences for either $x\to z$ or $y\to z$, is applied. 
Taking the limit $\omega_{1,2}\to 0$ we then obtain the result
$$
R =  - {\ts {1\over 12}} (5B+2C+6D) = {\ts {1\over 6}}(I+J)\, , 
\eqno (4.16) $$
where $E$ has been eliminated as a consequence of imposing (4.7).

The additional freedom present in this three point function beyond a minimal
expression satisfying the Ward identities is associated with the
arbitrariness in (1.3) since this allows a trivial solution of the Ward
identities if the three point function of the form
$$
\Gamma^{FFT}_{\mu\si,\nu\rho,\alpha\beta}(x,y,z)_C =
\pr^z_{\, \gamma} \pr^z_{\, \delta} 
\Gamma^{FFC}_{\mu\si,\nu\rho,\alpha\gamma\delta\beta}(x,y,z) \, ,
\eqno (4.17) $$
where $\Gamma^{FFC}$ has the Weyl tensor symmetries with respect to the
indices $\alpha\gamma\delta\beta$ exhibited in (1.4). 
$\Gamma^{FFC}_{\mu\si,\nu\rho,\alpha\gamma\delta\beta}(x,y,z)$
may be found according to the general rules for obtaining conformal
invariant three point functions. In general the resulting expression
is less singular for $x\sim y\sim z$
and hence there is no pole for $d\to 4$, such as in (4.13).
A particular solution may be found by taking
$$
\Gamma^{FFC}_{\mu\si,\nu\rho,\alpha\gamma\delta\beta}(x,y,z) =
{\I^F{}_{\!\!\mu\si,\mu'\si'}(x-z)\I^F{}_{\!\!\nu\rho,\nu'\rho'}(y-z)
\over \big ((x-z)^2(y-z)^2\big )^{d-2}}\, \E^C{}_{\!\! \mu'\si'\nu'\rho',
\alpha\gamma\delta\beta} {Q\over (Z^2)^{\hh d - 1}} \, .
\eqno (4.18) $$
Here $\E^C$ represents the projector onto tensors with Weyl symmetry and
is defined in detail in appendix A. The expression for the three point function
obtained from (4.17,18) can be related to the previous forms given by
(4.1,2,3) most directly by considering
$$
{\tilde t}^{FFT}_{\,\mu\si,\nu\rho,\alpha\beta} (X)_C = \pr_\gamma\pr_\delta
\Big ( \I^F{}_{\!\!\nu\rho,\nu'\rho'}(X) \, \E^C{}_{\!\! \mu\si\nu'\rho',
\alpha\gamma\delta\beta} {Q\over (X^2)^{\hh d - 1}} \Big ) \, .
\eqno (4.19) $$
Comparing the result of this calculation with (4.5) we may then find the
coefficients $A,B,C,D,E$ in terms of $Q$, or equivalently
$$
J = - I = {d^2(d-3)\over 2(d-1)}Q \, , \quad K=E = 0 \, , \quad
D = - {d^2\over (d-1)(d-2)}Q \, .
\eqno (4.20) $$
The result of using (4.17,18) in (4.1) gives an expression for 
$\l V_\mu (x) V_\nu(y) T_{\alpha\beta} (z) \r$ in which the conditions
on the energy momentum tensor in (1.1) are trivially satisfied.

To find a particular form for $\l V_\mu (x) V_\nu(y) T_{\alpha\beta} (z) \r$
in which the Ward identities are non trivial we consider the simple case
of taking in (4.3) $A=B=C=0, \, E=-4D$, which satisfies (4.7). With this choice
then using (2.9) we find the relatively simple expression
$$ \eqalign { \!\!\!\!\!\!\!\!
\Gamma^{FFT}_{\mu\si,\nu\rho,\alpha\beta}(x,y,z)_D & =  D \,
{\I^F{}_{\!\!\mu\si,\nu\rho}(x-y)\over \big ((x-y)^2 \big )^{d-2}}\,
(Z^2)^{\hh d}
\Big ( {Z_\alpha Z_\beta \over Z^2} - {1\over d} \de_{\alpha\beta} \Big ) \cr
& \ {} + \mu^{-\vep} {D\over 2\vep} \big (
\E^F{}_{\!\! \mu\si,\lambda(\alpha} \E^F{}_{\!\! \nu\rho,\lambda | \beta)}
- \quar \de_{\alpha\beta}\E^F{}_{\!\!\mu\si,\nu\rho} \big ) 
S_d{}^{\! 2} \de^d(x-z) \de^d(y-z) \, , \cr}
\eqno (4.21) $$
where we have subtracted the pole in $\vep$, as calculated in general
in (4.13,16), in accordance with standard dimensional regularisation so as to
ensure a well defined distribution for $d\approx 4$ for this  
conformally covariant three point function. The form of the residue of
the $\vep$ pole in (4.21) has been modified by terms of $\rO(\vep)$ from
(4.13) in order later to ensure compatibility with the standard form of Ward
identities. As usual the counterterm in (4.21) contains an arbitrary scale
$\mu$. The dependence of the regularised expression on $\mu$ reflects the short
distance singularity present for $d=4$.

To obtain the Ward identities flowing from the conservation equation for the
energy momentum tensor we may use, for general $d$,
$$ \eqalign { \!\!\!\!\!\!\!
\pr^z{}_{\!\! \alpha} \bigg ( (Z^2)^{\hh d}
\Big ( {Z_\alpha Z_\beta \over Z^2} - {1\over d} \de_{\alpha\beta} \Big )\bigg )
= - S_d {d-1\over d^2}& \Big ( \pr_\beta \de^d(z-x) + \pr_\beta \de^d(z-y) \cr
& + 2d \big ( \de^d(z-x) - \de^d(z-y) \big ) {(x-y)_\beta\over (x-y)^2 }\Big ) 
\, , \cr}
\eqno (4.22) $$
where the result is unambiguous if the traceless condition is imposed on the
distribution formed from the $Z$-dependent part of (4.21). If we use (4.22)
in conjunction with
$$
\pr^y{}_{\!\! \rho} \, 
{\I^F{}_{\!\!\mu\si,\nu\rho}(x-y)\over (x-y)^{2(d-2)}} = 0 \, ,
\quad \pr^y{}_{\!\! \rho} \, {(x-y)_\beta\over (x-y)^2} = - {I_{\beta\rho}
(x-y) \over (x-y)^2} \, ,
\eqno (4.23) $$
then we find
$$ \eqalign { \!\!\!\!
\pr^x_{\, \si} \pr^y_{\, \rho} \pr^z{}_{\!\! \alpha} 
\Gamma^{FFT}_{\mu\si,\nu\rho,\alpha\beta}(x,y,z)_D 
= {}& \pr^x_{\, \si} \Big ( \de^d(x-z) \big ( \de_{\si\beta} \Ga_{\mu\nu}
(x-y)  - \de_{\mu\beta} \Ga_{\si\nu}(x-y) \big )\Big ) \cr 
& + x ,\, \mu \leftrightarrow y,\,  \nu \, . \cr}
\eqno (4.24) $$
where we take
$$
\Ga_{\mu\nu}(x-y)=  DS_d {d-1\over d} \, {I_{\mu\nu}(x-y)\over(x-y)^{2(d-1)}}
+ \mu^{-\vep} {D\over 16\vep}S_d{}^{\! 2}
(\pr_\mu\pr_\nu - \de_{\mu\nu} \pr^2 ) \de^d(x-y) \, .
\eqno (4.25) $$
It should be noted that the $\pr \de^d$ terms in (4.22) do not contribute
to the final result in (4.24). In order to obtain (4.24), where the r.h.s. is
expressed solely in terms of $\Gamma_{\mu\nu}$, it is crucial that 
the second term in (4.21) has exactly the tensorial form shown. Since, by
extension of results such as (4.9), 
$$
{ I_{\mu\nu}(x) \over x^{2(d-1)}} \sim - {1\over 12\vep} S_d
(\pr_\mu\pr_\nu - \de_{\mu\nu} \pr^2 ) \de^d(x) \ \ \hbox{for} \ \ \vep\to 0\, ,
\eqno (4.26) $$
then it is easy to see that $\Ga_{\mu\nu}$ in (4.25) has the appropriate
dimensionally regularised form for $d\approx 4$. The result (4.24) has
the correct form expected for the Ward identity, if we take in (2.30)
$(C_{\beta\lambda})_{\mu\mu'}= - \E^T{}_{\!\!\beta\mu,\lambda\mu'}$, 
$$\eqalign {\!\!
\pr^z{}_{\!\! \alpha} \l V_\mu (x) V_\nu(y) T_{\alpha\beta} (z) \r ={}&
- \pr^x{}_{\!\! \si}\Big ( \de^d(x-z) \big ( \de_{\si\beta}
\l V_\mu(x) V_\nu(y) \r - \de_{\mu\beta} \l V_\si(x) V_\nu(y)\r\big )\Big ) \cr
& + x , \,  \mu \leftrightarrow y,\, \nu \, , \cr}
\eqno (4.27) $$
assuming the two point function $ \l V_\mu (x) V_\nu(y)\r 
= - \Ga_{\mu\nu}(x-y)$ (which satisfies the conservation equation
$\pr^x{}_{\!\!\mu} \l V_\mu (x) V_\nu(y)\r = 0$).
The overall coefficient, defined as in (2.17) and generalising to the
general case as given by (4.12), is then
$$
C_V = {1\over d} S_d \, (I+J) \, .
\eqno (4.28) $$
Although the identity (4.27) is satisfied with this construction the
regularised expression (4.21) now has an anomaly in the trace of the
energy momentum tensor arising from the counterterm since the $\vep$ pole
is cancelled by the operation of taking the trace,
$$
\Gamma^{FFT}_{\mu\si,\nu\rho,\alpha\alpha}(x,y,z)_D = \mu^{-\vep}
{\ts {1\over 8}} D\,
\E^F{}_{\!\!\mu\si,\nu\rho} S_d{}^{\! 2} \de^d(x-z) \de^d(y-z) \, .
\eqno (4.29) $$
Hence in four dimensions the trace anomaly in the three point function
for the energy momentum tensor and two conserved currents becomes
$$
\l V_\mu (x) V_\nu(y) T_{\alpha\alpha} (z) \r = -{\ts {1\over 3}}\pi^2C_V
\E^F{}_{\!\!\mu\si,\nu\rho} 
\pr^x_{\, \si} \pr^y_{\, \rho}\big ( \de^4(x-z) \de^4(y-z) \big ) \, .
\eqno (4.30) $$

This result is as expected for the trace anomaly in four dimensions (note
that as $C_{\alpha\alpha}=0$ in this case there is no contribution of the
two point function $\l V V \r$ to the trace identity). If
the conserved currents $V_\mu$ are coupled to a background gauge potential
$A_\mu$ then the expectation value for the energy momentum tensor
trace in this background has contributions proportional to $F_{\mu\nu}
F_{\mu\nu}$ as shown in (1.7). By considering functional
derivatives with respect to $A_\mu$ it is easy to see that the trace
anomaly is exactly compatible with (4.30) if $\kappa$ is given by (2.46).
\bigskip
\leftline{\bigbf 5 Energy Momentum Tensor Three Point Function}
\medskip
A significant result obtained in \one\ is that there are only three linearly
independent conformally
covariant forms for the conserved and traceless energy momentum tensor 
symmetric three point function in general dimensions $d$ (although for $d=2$
there is only one while if $d=3$ there are two).
We here rederive this result more simply by following an approach 
in which the Bose symmetry of the three point function is always manifest.
It is convenient to define
$$  
\l T_{\mu\nu} (x) T_{\si\rho}(y) T_{\alpha\beta} (z) \r =  
\Ga^{TTT}_{\, \mu\nu,\si\rho,\alpha\beta}(x,y,z) \, , 
\eqno (5.1) $$ 
and then using the results on conformal transformations 
in section 2, in particular the existence  of the vectors $X,Y,Z$ which
are given by (2.7) and its obvious permutations, it is possible to construct
five possible completely symmetric expressions, 
assuming $T_{\mu\nu}$ is a traceless tensor field of dimension $d$, for
$\Ga^{TTT}_{\, \mu\nu,\si\rho,\alpha\beta}(x,y,z)$ with the required properties
under conformal transformations so that in general we may write an expansion
involving just five coefficients $\A,\B,\C,\D,\E$,
$$ \eqalignno {
\big ( & (x-y)^2(y-z)^2(z-x)^2 \big )^{\hh d}\,
\Ga^{TTT}_{\, \mu\nu,\si\rho,\alpha\beta}(x,y,z) \cr
&= \E^T{}_{\!\!\mu\nu,\mu'\nu'} \E^T{}_{\!\!\si\rho,\si'\rho'}
\E^T{}_{\!\!\alpha\beta,\alpha'\beta'} 
\Big \{ \A\, I_{\nu'\si'}(x-y) I_{\rho'\alpha'}
(y-z) I_{\beta'\mu'}(z-x) \cr
&\qquad \qquad \qquad \qquad + \B\, I_{\mu'\si'}(x-y) I_{\nu'\alpha'}(x-z) 
Y_{\rho'}Z_{\beta'} (y-z)^2 + \hbox{cyclic permutations} \Big \} \cr
&\  + \C \, \I^T{}_{\!\!\mu\nu,\si\rho}(x-y)
\Big ({Z_\alpha Z_\beta\over Z^2} - {1\over d}\de_{\alpha\beta} \Big ) 
+ \hbox{cyclic permutations} \cr
&\ + \D \, \E^T{}_{\!\!\mu\nu,\mu'\nu'} \E^T{}_{\!\!\si\rho,\si'\rho'}
X_{\mu'}Y_{\si'} (x-y)^2 I_{\nu'\rho'}(x-y)
\Big ({Z_\alpha Z_\beta\over Z^2} - {1\over d}\de_{\alpha\beta} \Big ) 
+ \hbox{cyclic permutations} \cr
&\ +\E \, \Big ({X_\mu X_\nu\over X^2} - {1\over d}\de_{\mu\nu} \Big )
\Big ({Y_\si Y_\rho\over Y^2} - {1\over d}\de_{\si\rho} \Big )
\Big ({Z_\alpha Z_\beta\over Z^2} - {1\over d}\de_{\alpha\beta} \Big ) \, . 
& (5.2) \cr}
$$
If all points are collinear the r.h.s. of (5.2) reduces to constant tensors
invariant under $O(d-1)$ preserving the line defined by $x,y,z$ and from
the results in \one\ it is easy to verify the completeness of the expansion
given by (5.2).

It remains only to impose the conservation equation
$$
\pr^x{}_{\!\!\mu} \l T_{\mu\nu} (x) T_{\si\rho}(y) T_{\alpha\beta} (z) \r
= 0 \ \ \hbox{for non coincident points} \, ,
\eqno (5.3) $$
which leads to relations between the coefficients $\A,\B,\C,\D,\E$.
The calculation of  the divergence
$\pr^x{}_{\!\!\mu}\Ga^{TTT}_{\, \mu\nu,\si\rho,\alpha\beta}(x,y,z)$, for
the general form given in (5.2), may be
simplified by identifying separately all contributions which preserve
manifest conformal covariance. To this end it is important to
recognise that the derivative $\pr^x{}_{\!\!\mu}$ acting on an expression which
is a scalar of scale dimension zero at $x$ gives a vector of scale 
dimension one under conformal transformations at $x$. As an illustration
it follows trivially from the definition of $Z$, and hence $Y$, in (2.7)
$$
\pr^x{}_{\!\!\mu} Y_\si = - {I_{\mu\si}(x-y) \over (x-y)^2} \, , \quad
\pr^x{}_{\!\!\mu} Z_\alpha = {I_{\mu\alpha}(x-z) \over (x-z)^2} \, .
\eqno (5.4) $$
For application in (5.3) this may be extended, by using also
$\pr^x{}_{\!\!\mu} Y^2 =  2X_\mu Y^2$, to give as well
$$
\pr^x{}_{\!\!\mu} \Big ({Y_\si Y_\rho\over Y^2} - {1\over d}\de_{\si\rho}\Big )
= -2{(y-z)^2\over (x-z)^2}\E^T{}_{\!\!\si\rho,\si'\rho'} I_{\mu\si'}(x-y)
Y_{\rho'} - 2 X_\mu 
\Big ({Y_\si Y_\rho\over Y^2} - {1\over d}\de_{\si\rho}\Big ) \, .
\eqno (5.5) $$
Furthermore the divergence of a symmetric traceless tensor of dimension $d$
is a vector.  Thus we may obtain
$$ \eqalign {
\pr^x{}_{\!\!\mu} \bigg ( & {1\over \big ( (x-y)^2(x-z)^2 \big )^{\hh d} }\, 
\E^T{}_{\!\!\mu\nu,\mu'\nu'} I_{\mu'\si}(x-y) I_{\nu'\alpha}(x-z) \bigg ) \cr
& = {d^2-4\over 2d} {(y-z)^2 \over \big ( (x-y)^2(x-z)^2 \big )^{\hh d} }
\Big ( {I_{\nu\alpha}(x-z) \over (x-z)^2} Y_\si
- {I_{\nu\si}(x-y) \over (x-y)^2} Z_\alpha \Big ) \, , \cr
\pr^x{}_{\!\!\mu} \bigg ( & {1\over \big ( (x-y)^2 \big)^{\hh d}
\big ((x-z)^2 \big )^{\hh d - 1} } \,
\E^T{}_{\!\!\mu\nu,\mu'\nu'} X_{\mu'} I_{\nu'\si}(x-y) \bigg ) \cr
& = {d-2\over 2d} {(y-z)^2 \over \big ( (x-y)^2(x-z)^2 \big )^{\hh d} }
\Big ( d\, X_\nu Y_\si - {I_{\nu\si}(x-y) \over (x-y)^2}  \Big ) \, , \cr
\pr^x{}_{\!\!\mu} \bigg ( & {1\over \big ( (x-y)^2(x-z)^2 \big )^{\hh d} } \,
\I^T{}_{\!\!\mu\nu,\si\rho}(x-y) \bigg ) \cr
& = d \, {(y-z)^2 \over \big ( (x-y)^2\big )^{\hh d}
\big ( (x-z)^2 \big )^{\hh d+1} }
\,  \E^T{}_{\!\!\si\rho,\si'\rho'} I_{\nu\si'}(x-y) Y_{\rho'}
\, , \cr
\pr^x{}_{\!\!\mu} \bigg ( & {1\over \big ( (x-y)^2(x-z)^2 \big )^{\hh d} } \,
\Big ({X_\mu X_\nu\over X^2} - {1\over d}\de_{\mu\nu} \Big )\bigg ) = 0 \, .\cr}
\eqno (5.6) $$
Using these results, together with relations of the form (2.9),
$\pr^x{}_{\!\!\mu} \Ga^{TTT}_{\, \mu\nu,\si\rho,\alpha\beta}(x,y,z)$
can also be obtained as a sum of conformally covariant forms. There are then
two conditions necessary to ensure (5.3)
$$ \eqalign {
R_1 \equiv {}& (d^2- 4) \A + (d+2) \B - 4d\, \C - 2\D =  0 \, , \cr
R_2 \equiv {}& (d-2)(d+4) \B - 2d(d+2)\C + 8 \D - 4 \E =  0 \, . \cr}
\eqno (5.7) $$
Thus there remain three independent coefficients which may be taken to
be $\A,\B,\C$ by using (5.7) to eliminate $\D,\E$.

For comparison with the general form (2.20) we may note that the
expression (5.1,2) can be rewritten identically as
$$
\l T_{\mu\nu} (x) T_{\si\rho}(y) T_{\alpha\beta} (z) \r =
{\I^T{}_{\!\!\mu\nu,\mu'\nu'}(x-z) \I^T{}_{\!\!\si\rho,\si'\rho'}(y-z)
\over \big ( (x-z)^2 (y-z)^2 \big )^d } \, t_{\mu'\nu',\si'\rho',\alpha\beta}
(Z) \, ,
\eqno (5.8) $$
where, from (5.2), we may obtain
$$ \eqalign { \!\!\!\!\!\!
t_{\mu\nu,\si\rho,\alpha\beta}(Z) = {}&  \A \,
\E^T{}_{\!\!\mu\nu,\ep\eta} \E^T{}_{\!\!\si\rho,\eta\lambda}
\E^T{}_{\!\!\alpha\beta,\lambda\ep} \, {1\over (Z^2)^{\hh d}} \cr
& + (\B - 2\A )\, \E^T{}_{\!\!\alpha\beta,\ep\eta} 
\E^T{}_{\!\!\si\rho,\eta\kappa} \E^T{}_{\!\!\mu\nu,\lambda\ep}
\, {Z_\kappa Z_\lambda \over (Z^2)^{\hh d + 1 }} \cr
& - \B \big ( \E^T{}_{\!\!\mu\nu,\ep\eta} \E^T{}_{\!\!\si\rho,\eta\kappa}
\E^T{}_{\!\!\alpha\beta,\lambda\ep} + (\mu\nu) \leftrightarrow (\si\rho)
\big ) {Z_\kappa Z_\lambda \over (Z^2)^{\hh d + 1 }} \cr
& + \C \bigg ( \E^T{}_{\!\!\mu\nu,\si\rho}
\Big ({Z_\alpha Z_\beta\over Z^2} - {1\over d}\de_{\alpha\beta} \Big ) \cr
& \qquad +  \E^T{}_{\!\!\si\rho,\alpha\beta}
\Big ({Z_\mu Z_\nu\over Z^2} - {1\over d}\de_{\mu\nu} \Big )
+ \E^T{}_{\!\!\alpha\beta,\mu\nu}
\Big ({Z_\si Z_\rho\over Z^2} - {1\over d}\de_{\si\rho} \Big ) \bigg )
{1\over (Z^2)^{\hh d}} \cr
& + (\D-4\C)\, \E^T{}_{\!\!\mu\nu,\ep\kappa}\E^T{}_{\!\!\si\rho,\ep\lambda}
\Big ({Z_\alpha Z_\beta\over Z^2} - {1\over d}\de_{\alpha\beta} \Big )
{Z_\kappa Z_\lambda \over (Z^2)^{\hh d + 1 }} \cr
& - (\D -2\B) \bigg (  \E^T{}_{\!\!\si\rho,\ep\kappa}
\E^T{}_{\!\!\alpha\beta,\ep\lambda}
\Big ({Z_\mu Z_\nu\over Z^2} - {1\over d}\de_{\mu\nu} \Big )
+ (\mu\nu) \leftrightarrow (\si\rho) \bigg )
{Z_\kappa Z_\lambda \over (Z^2)^{\hh d + 1 }} \cr
& + (\E +4\C- 2\D) \Big ({Z_\mu Z_\nu\over Z^2} - {1\over d}\de_{\mu\nu} \Big )
\Big ({Z_\si Z_\rho\over Z^2} - {1\over d}\de_{\si\rho} \Big )
\Big ({Z_\alpha Z_\beta\over Z^2} - {1\over d}\de_{\alpha\beta} \Big )
{1\over (Z^2)^{\hh d}} \, . 
\cr} 
\eqno (5.9)$$
which satisfies the conservation equation, 
$\pr_\mu t_{\mu\nu,\si\rho,\alpha\beta}(Z) = 0$, subject to (5.7), which
can alternatively be derived directly from (5.8), and also
$ \I^T{}_{\!\!\mu\nu,\mu'\nu'}(Z) t_{\alpha\beta,\mu'\nu',\si\rho}(Z)
= t_{\mu\nu,\si\rho,\alpha\beta}(Z)$, arising from applying the symmetry 
condition (2.24) to (5.8)\footnote{*}{In terms of the treatment in ref.(\one), 
where the three point function was specified in terms of coefficients $a,b,c$
or alternatively $r,s,t$, the relationship is $\A=8a=8t$, 
$\B=8(b+2a)= - 4(dr+(d+4)s)$, $\C=2c=-2(ds+4t)$}. If we use (4.9),
and its generalisations, to determine the short distance singularity similarly
to (4.10) we find
$$
x^{2 \omega}t_{\mu\nu,\si\rho,\alpha\beta}(x) \sim {1\over 2\omega} \,
{(d+4)R_1 - 2R_2 \over d(d+2)(d+4)} \,
\E^T{}_{\!\!\mu\nu,\ep\eta} \E^T{}_{\!\!\si\rho,\eta\lambda}
\E^T{}_{\!\!\alpha\beta,\lambda\ep}\, S_d \de^d(x) \, , 
\eqno (5.10) $$
with $R_1 , R_2$ given by (5.7). The pole at $\omega=0$ is therefore absent
when the conservation equations (5.7) are imposed. 

If we use the expression (5.9) it is straightforward to verify that
$$ \eqalign{
\int \!\! \d\Omega_{\hx} \, \hx_\mu \hx_\nu 
t_{\mu\nu,\si\rho,\alpha\beta}(\hx) = {}& -d C_T \,
\E^T{}_{\!\!\si\rho,\alpha\beta} \, , \cr
\int \!\! \d\Omega_{\hx} \, \hx_\mu \hx_{[\omega}
t_{\mu|\nu],\si\rho,\alpha\beta}(\hx) = {}& - 2C_T \,
\E^T{}_{\!\!\si\rho,\lambda[\omega} \E^T{}_{\!\!\nu]\lambda,\alpha\beta}\, ,\cr}
\eqno (5.11) $$ 
where, employing (5.7),
$$
C_T = {S_d \over d(d+2)}\big ( \half (d+2)(d-1) \A - \B - 2(d+1) \C \big ) \, .
\eqno (5.12) $$
The result (5.11) in exact agreement with the form dictated by the conformal
identities (2.29) with $C_T$ the coefficient of the energy  momentum tensor
two point function as shown in (2.18).

Although the expression given by (5.2), subject to (5.7), is conformally
covariant and satisfies the Ward identities (2.34,35) for general $d$ it
has singularities for $x,y,z$ coincident when $d=4$. In terms of dimensional
regularisation there are two possible counterterms which are 
compatible with Bose symmetry and other identities so that we may take
$$ \eqalign{
\l T_{\mu\nu} (x) T_{\si\rho}(y) T_{\alpha\beta} (z) \r = {}&
\Ga^{TTT}_{\, \mu\nu,\si\rho,\alpha\beta}(x,y,z)  \cr
& - 8 {\mu^{-\vep}\over \vep}
\big ( \beta_a D^F_{\, \mu\nu,\si\rho,\alpha\beta}(x,y,z) +
\beta_b D^G_{\, \mu\nu,\si\rho,\alpha\beta}(x,y,z) \big ) \, , \cr}
\eqno (5.13) $$
where $D^F,D^G$ should be proportional to $\de^d(x-y)\de^d(x-z)$ with four
derivatives. The simplest method of defining $D^F,D^G$ is in terms of 
functional derivatives of the possible local counterterms for a curved space
background metric which are formed from the metric $g_{\mu\nu}$,
$$ \eqalign{
D^F_{\, \mu\nu,\si\rho,\alpha\beta}(x,y,z) = {}&
{\de^3\over \de g^{\mu\nu}(x)g^{\si\rho}(y)g^{\alpha\beta}(z)}
\int \! \d^d x \, \sqrt g F  \Big |_{\rm{flat \ space}} \, , \cr
D^G_{\, \mu\nu,\si\rho,\alpha\beta}(x,y,z) = {}&
{\de^3\over \de g^{\mu\nu}(x)g^{\si\rho}(y)g^{\alpha\beta}(z)}
\int \! \d^d x \, \sqrt g G  \Big |_{\rm{flat \ space}} \, , \cr}
\eqno (5.14) $$
with now, for general $d$, replacing (1.8),
$$
F = C^{\gamma\de\chi\omega} C_{\gamma\de\chi\omega} \, , \quad
G =  6\, R^{\gamma\de}{}_{[\gamma\de} R^{\chi\omega}{}_{\chi\omega]}
= R^{\gamma\de\chi\omega} R_{\gamma\de\chi\omega} - 4\,  R^{\gamma\de}
R_{\gamma\de} + R^2 \, .
\eqno (5.15) $$
Using diffeomorphism invariance, since $F,G$ are scalars, it is not difficult
to see that the expressions defined by (5.14) satisfy identities such that
(5.13) is compatible with the Ward identity (2.34) if we now take for the 
two point function
$$
\l T_{\si\rho} (y) \, T_{\alpha \beta} (z) \r = C_T \,
{\I^T{}_{\!\! \si \rho,\alpha\beta}(y-z)\over (y-z)^{2d}} + 16\beta_a
{\mu^{-\vep}\over \vep} 
\E^C{}_{\!\! \si\ep\eta\rho,\alpha\gamma\de\beta}\, \pr^y{}_{\!\!\ep}
\pr^y{}_{\!\!\eta}
\pr^z{}_{\!\!\gamma} \pr^z{}_{\!\!\de} \de^d(y-z)  \, . 
\eqno (5.16) $$
In principle $\beta_a,\beta_b$ should be determined in terms of $\A,\B,\C$
by requiring the (5.13) gives a well defined regularised expression for
$d\to 4$ but the appropriate extensions of formulae such as (4.14) are too
involved to allow for simple analysis. In the next section we obtain a
result for $\beta_b$ by a less direct route and it is clear from (2.37) that
requiring (5.16) is the dimensionally regularised two point function leads to
(2.42) again where, from (5.12) with $d=4$, we have
$C_T={1\over 12}\pi^2(9\A-\B-10\C)$. From its definition in (5.14)
$$ \eqalign {
D^F_{\, \mu\mu,\si\rho,\alpha\beta}(x,y,z)={}& 
- \big(\de^d(x-y)+\de^d(x-z)\big) {\de^2\over g^{\si\rho}(y)g^{\alpha\beta}(z)}
\int \! \d^d x \, \sqrt g F  \Big |_{\rm{flat \ space}} \cr
& + \half \vep \, {\de^2\over g^{\si\rho}(y)g^{\alpha\beta}(z)} F(x)
\Big |_{\rm{flat \ space}} \cr}
\eqno (5.17) $$
and hence, using (A.4),
$$ \eqalign {
D^F_{\, \mu\mu,\si\rho,\alpha\beta}(x,y,z)={}& 
- 8 \big(\de^d(x-y)+\de^d(x-z)\big)
\E^C{}_{\!\! \si\ep\eta\rho,\alpha\gamma\de\beta}\pr_\ep\pr_\eta
\pr_\gamma\pr_\de \de^d(y-z) \cr
& +\vep \, 4 \E^C{}_{\!\! \si\ep\eta\rho,\alpha\gamma\de\beta}\,
\pr_{\ep} \pr_{\eta} \de^d(x-y) \pr_{\gamma} \pr_{\de} \de^d(x-z) \, ,\cr}
\eqno (5.18) $$
and a corresponding identity for $D^G$ given later, it not difficult to see
that the counterterms in (5.14) generate exactly the finite trace anomalies
exhibited in (1.9) with the required coefficients $\beta_a,\beta_b$. 

The coefficients $\A,\B,\C$ and hence via (5.7) $\D,\E$ should be calculable
in any conformal field theory. In four dimensions there are three basically
trivial theories given by free scalars, spin $\half$ fermions and spin $1$
vectors for which \one\
$$ \eqalign{
\A ={}&  {1\over \pi^6} \Big ( {\ts {8\over 27}} n_S - 16 n_V \Big ) \, , \cr
\B ={}&  {- {1\over \pi^6}} \Big ( {\ts {16\over 27}} n_S + 4 n_F + 32 n_V
\Big ) \, , \cr
\C ={}&  {- {1\over \pi^6}} \Big ( {\ts {2 \over 27}} n_S + 2 n_F + 16 n_V
\Big ) \, ,  \cr}
\eqno (5.19) $$
where $n_S, \, n_F , \, n_V$ are the number of free scalar, Dirac fermion
and vector fields. Substituting in (5.12) gives the standard results for
$C_T$ in free field theories. For general $d$ only free scalars or fermions
give conformal theories, the results are given in appendix B.
\bigskip
\leftline{\bigbf 6 Use of Derivative Forms for the Energy Momentum Tensor}
\medskip
In order to determine explicitly the coefficients of the local
singularities in the energy momentum tensor three point function
 when $x,y,z$ become coincident, and hence obtain results for the coefficients
in the trace anomaly, it is much simpler if we make use of
the potential freedom expressed in (1.3) to reduce the degree of the
singularity by extracting derivatives. To this end we consider a three
point function which is restricted to the form
$$
\Ga^{TTT}_{\, \mu\nu,\si\rho,\alpha\beta}(x,y,z)_C
= \pr^x_{\,\kappa}\pr^x_{\,\lambda} \pr^y_{\,\ep}\pr^y_{\,\eta}
\Gamma^{CCT}_{\mu\kappa\lambda\nu,\si\ep\eta\rho,\alpha\beta}(x,y,z) \, ,
\eqno (6.1) $$
where $\Gamma^{CCT}_{\mu\kappa\lambda\nu,\si\ep\eta\rho,\alpha\beta}$ has
the Weyl tensor symmetries (1.4) with respect to the indices
$\mu\kappa\lambda\nu$ and $\si\ep\eta\rho$. By virtue of the results in
section 2 it is still feasible to express
$\Gamma^{CCT}_{I,J,\alpha\beta}$, where $I,J$ are shorthand notation
for multi-indices with symmetry (1.4), in the conformal covariant form
$$
\Gamma^{CCT}_{I,J,\alpha\beta}(x,y,z)
= {\I^C{}_{\!\!I,I'}(x-z)\I^C{}_{\!\! J,J'}(y-z)
\over \big ((x-z)^2(y-z)^2\big )^{d-2}}\,
t^{CCT}_{\, I',J', \alpha\beta}(Z) \, .
\eqno (6.2) $$ 
The general expression for $t^{CCT}$ has ten terms
$$ \eqalign{\!\!\!\!
t^{CCT}_{\mu\kappa\lambda\nu,\si\ep\eta\rho,\alpha\beta}(Z) {}&  =
A\, \E^C{}_{\!\!\mu\kappa\lambda\nu,\alpha'\tau\chi\omega}
\E^C{}_{\!\!\si\ep\eta\rho,\beta'\tau\chi\omega}
\E^T{}_{\!\!\alpha'\beta',\alpha\beta}\, {1\over (Z^2)^{\hh d - 2}} \cr
&\ \  + B\, \E^C{}_{\!\!\mu\kappa\lambda\nu,\si\ep\eta\rho}
\Big ( {Z_\alpha Z_\beta \over Z^2}
- {1\over d} \de_{\alpha\beta} \Big ) {1\over (Z^2)^{\hh d - 2}} \cr
&\ \  + C \,\E^C{}_{\!\!\mu\kappa\lambda\nu,\alpha'\theta\chi\omega}
\E^C{}_{\!\!\si\ep\eta\rho,\beta'\phi\chi\omega}
\E^T{}_{\!\!\alpha'\beta',\alpha\beta}\, {Z_\theta Z_\phi\over
(Z^2)^{\hh d - 1}} \cr
&\ \  + C'\, \E^C{}_{\!\!\mu\kappa\lambda\nu,\alpha'\chi\omega\theta}
\E^C{}_{\!\!\si\ep\eta\rho,\beta'\chi\omega\phi}
\E^T{}_{\!\!\alpha'\beta',\alpha\beta}\, {Z_\theta Z_\phi\over
(Z^2)^{\hh d - 1}} \cr
&\ \  + D\, \big ( \E^C{}_{\!\!\mu\kappa\lambda\nu,\chi \alpha\beta \omega}
\E^C{}_{\!\!\si\ep\eta\rho,\chi\theta\phi\omega} +
\E^C{}_{\!\!\mu\kappa\lambda\nu,\chi\theta\phi\omega}
\E^C{}_{\!\!\si\ep\eta\rho,\chi \alpha\beta \omega} \big )
{Z_\theta Z_\phi\over (Z^2)^{\hh d - 1}} \cr
&\ \  + E\, \big ( \E^C{}_{\!\!\mu\kappa\lambda\nu,\alpha'\tau\chi\omega}
\E^C{}_{\!\!\si\ep\eta\rho,\theta\tau\chi\omega}\cr
&\qquad \qquad + \E^C{}_{\!\!\mu\kappa\lambda\nu,\theta\tau\chi\omega}
\E^C{}_{\!\!\si\ep\eta\rho,\alpha'\tau\chi\omega}\big)
\E^T{}_{\!\!\alpha'\beta',\alpha\beta}\, {Z_\theta Z_{\beta'}\over
(Z^2)^{\hh d - 1}} \cr
&\ \ + F \, \E^C{}_{\!\!\mu\kappa\lambda\nu,\alpha'\theta\phi\omega}
\E^C{}_{\!\!\si\ep\eta\rho,\beta'\theta'\phi'\omega}
\E^T{}_{\!\!\alpha'\beta',\alpha\beta}\, {Z_\theta Z_\phi
Z_{\theta'}Z_{\phi'} \over(Z^2)^{\hh d}} \cr
&\ \ + G \, \E^C{}_{\!\!\mu\kappa\lambda\nu,\theta\tau\chi\omega}
\E^C{}_{\!\!\si\ep\eta\rho,\phi\tau\chi\omega}
\Big ( {Z_\alpha Z_\beta \over Z^2} - {1\over d} \de_{\alpha\beta} \Big )
{Z_\theta Z_\phi\over (Z^2)^{\hh d }} \cr
&\ \  + H\, \big ( \E^C{}_{\!\!\mu\kappa\lambda\nu,\alpha'\tau\chi\theta'}
\E^C{}_{\!\!\si\ep\eta\rho,\theta\tau\chi\phi}\cr
&\qquad \qquad + \E^C{}_{\!\!\mu\kappa\lambda\nu,\theta\tau\chi\phi}
\E^C{}_{\!\!\si\ep\eta\rho,\alpha'\tau\chi\theta'}\big)
\E^T{}_{\!\!\alpha'\beta',\alpha\beta}\, {Z_\theta Z_\phi Z_{\theta'}
Z_{\beta'}\over (Z^2)^{\hh d - 1}} \cr
&\ \ + I \, \E^C{}_{\!\!\mu\kappa\lambda\nu,\theta\tau\chi\phi}
\E^C{}_{\!\!\si\ep\eta\rho,\theta'\tau\chi\phi'}
\Big ( {Z_\alpha Z_\beta \over Z^2} - {1\over d} \de_{\alpha\beta} \Big )
{Z_\theta Z_\phi Z_{\theta'}Z_{\phi'} \over(Z^2)^{\hh d}} \, . \cr}
\eqno (6.3) $$
When $d=4$ this set is overcomplete since there are various identities
which arise from the vanishing of tensors antisymmetric on five indices.
Thus, if $C_{\mu\si\rho\nu}$ satisfies (1.4),
$$ \eqalign{
30\,C_{\alpha\kappa[\lambda\theta} X_\phi C_{\beta\kappa]\lambda\phi} X_\theta
= {}& \big ( 2 C_{\alpha\kappa\lambda\theta}C_{\beta\kappa\lambda\phi}
+ C_{\alpha\theta\kappa\lambda}C_{\beta\phi\kappa\lambda}
- 2 C_{\kappa\alpha\beta\lambda} C_{\kappa\theta\phi\lambda} \big )
X_\theta X_ \phi \cr
& - C_{\alpha\kappa\lambda\omega}C_{\beta\kappa\lambda\omega} X^2
+ C_{\alpha\theta\kappa\omega} C_{\theta\kappa\lambda\omega} 
X_\theta X_\beta \, , \cr
30\,C_{\ep\eta[\kappa\lambda} X_\alpha C_{\ep\eta]\kappa\lambda}
= {}& C_{\ep\eta\kappa\lambda} C_{\ep\eta\kappa\lambda} X_\alpha - 4
C_{\alpha\eta\kappa\lambda} C_{\beta\eta\kappa\lambda} X_\beta \, . \cr}
\eqno (6.4) $$
Hence for $d=4$ it is easy to see that for instance
$\E^C{}_{\!\!\mu\kappa\lambda\nu,\alpha'\tau\chi\omega}
\E^C{}_{\!\!\si\ep\eta\rho,\beta'\tau\chi\omega}
\E^T{}_{\!\!\alpha'\beta',\alpha\beta}=0$ and thus $t^{CCT}$ is independent
of $A$. In general it depends only on
$$
4B+2E+G+D\, , \quad C+D \, , \quad C'+2D\, , \quad F\, , \ H \, , \ I \, .
\eqno (6.5) $$

The expressions given by (6.1,2,3) can be related to the previous
general results for the energy momentum tensor three point function
since expressing (6.1) in the form (5.8) requires
$$
t_{\mu\nu,\si\rho,\alpha\beta}(Z)_C
= \pr_\kappa \pr_\lambda \pr_\ep \pr_\eta
t^{CCT}_{\mu\kappa\lambda\nu,\si\ep\eta\rho,\alpha\beta}(Z) \, .
\eqno (6.6) $$
By construction the result given by (6.6) automatically satisfies the
conservation equation $\pr_\mu t_{\mu\nu,\si\rho,\alpha\beta}(Z) = 0$ and
this in turn can be shown to also imply the necessary symmetry conditions
$t_{\mu\nu,\si\rho,\alpha\beta}(Z)
= \I^T{}_{\!\!\mu\nu,\mu'\nu'}(Z) t_{\alpha\beta,\mu'\nu',\si\rho}(Z)$,
if $t_{\mu\nu,\si\rho,\alpha\beta}(Z)=t_{\si\rho,\mu\nu,\alpha\beta}(Z)$,
which have already been imposed on the expression exhibited in (5.9). Hence
$t_{\mu\nu,\si\rho,\alpha\beta}(Z)_C$ gives results for the coefficients 
$\A,\B,\C,\D,\E$ satisfying (5.7) so it is sufficient to quote only $\A,\B,\C$.
The results for general $d$ are lengthy so they are relegated to appendix C
but for $d=4$ we have
$$ \eqalign {
\A ={}& {\ts {1\over 9}}\big ( 6(4B+2E+G+D) + 8(C+D) + 7(C'+2D) - \half F
+ 15H  + 12I \big )
\, , \cr
\B ={}& {\ts {1\over 9}}\big (24(4B+2E+G+D) + 2(C+D) + 13(C'+2D) + 
{\ts {11\over 2}}F + 15H + {\ts {51\over 2}} I\big )
\, , \cr
\C ={}& {\ts {1\over 9}}\big ( 3(4B+2E+G+D) + 7(C+D) + 5(C'+2D) - F +
12 H + {\ts {33\over 4}}I \big ) \, , \cr}
\eqno (6.7) $$
which are in accord with the requirement of depending only on the linear 
combinations appearing in (6.5).
It is important to note that substituting in (5.12), $C_T=0$, for general $d$,
so that assuming the form (6.1) gives only two linearly independent solutions
for the conformally covariant energy momentum tensor three point function.

Besides $t^{CCT}$ it is also natural to define
$$
{\tilde t}^{CCT}_{I,J,\alpha\beta}(X) = {1\over (X^2)^2}\I^C{}_{\!\! J,J'}(X)
t^{CCT}_{I,J',\alpha\beta}(-X) \, ,
\eqno (6.8) $$
so that in (6.2) the short distance limit as $y\to z$ is given by
$$
\Gamma^{CCT}_{I,J,\alpha\beta}(x,y,z) \sim
{\I^C{}_{\!\!I,I'}(x-z) \over (x-z)^{2(d-2)}}\,
{\tilde t}^{CCT}_{\, I',J, \alpha\beta}(y-z) \, .
\eqno (6.9) $$
If the conservation equation on $\Gamma^{CCT}$ is imposed then
$$
\pr^z_{\, \alpha}\Gamma^{CCT}_{I,J,\alpha\beta}(x,y,z) = 0 \ \ 
\Rightarrow \ \ \pr^X_{\, \alpha} {\tilde t}^{CCT}_{I,J,\alpha\beta}(X)
= 0 \, ,
\eqno (6.10) $$
which, after calculating ${\tilde t}^{CCT}$ from (6.3), leads to two
conditions,
$$ \eqalign{
T_1 \equiv {}& - (d^2-16)(4A+C') + (d+4)\big (4C + 2(d-2)D + F \big ) \cr
& + 16 (4B+2E + G + D) + 4I + 8 H =0 \, , \cr
T_2 \equiv {}& - (d^2-16)A -\half (d-4) C' + (d+2) C +3dD + 16B + 4E + 2G  =0
\, . \cr}
\eqno (6.11) $$
The second condition, $T_2=0$, is not necessary if $d=4$.
We may also calculate the short distance singularity in $\Gamma^{CCT}(x,y,z)$
as $y\to z$ or $x\to z$ by virtue of (6.9) by considering
$$
x^{2\omega} {\tilde t}^{CCT}_{I,J,\alpha\beta}(x) \sim {1\over 2\omega}\,
{U\over d(d+2)(d+4)} \, \E^C{}_{\!\!I,\alpha'\tau\chi\omega}
\E^C{}_{\!\!J,\beta'\tau\chi\omega}\E^T{}_{\!\!\alpha'\beta',\alpha\beta}
\, S_d \de^d(x) \, , 
\eqno (6.12) $$
where
$$ \eqalign{
U = {}& (d^2-16)\big ( (d-2)A - C-C'\big ) -16(d-2)B - 6dD \cr
& - 2(d-8)(G+2E) +{\ts {3\over 2}}(d+4)F + 6(I+2H) \cr
= {}& {\ts {3\over 2}} T_1 - (d+4)T_2 \, . \cr}
\eqno (6.13) $$
If $U=0$ there are no subdivergences in
$\Gamma^{CCT}$ for general $d$. It should be noted that if $d=4$ this
condition is superfluous since the tensorial expression in (6.12), as
remarked earlier, vanishes identically.

The reason for here  writing the three point function in the form (6.1) is that
the singular behaviour as $d\to 4$ may be quite straightforwardly 
determined since it involves
$\delta$-functions without derivatives. Imposing the condition that the
residue of the pole in $\vep$ should be traceless in $\alpha\beta$ we are
led to unique form
$$
\Gamma^{CCT}_{I,J,\alpha\beta}(x,y,z) \sim
{R\over \vep}\, \E^C{}_{\!\!I,\alpha\tau\chi\omega}
\E^C{}_{\!\!J,\beta'\tau\chi\omega} \E^T{}_{\!\!\alpha'\beta',\alpha\beta} \,
S_d{}^{\! 2} \de^d(x-z) \de^d(y-z) \, .
\eqno (6.14) $$
The tensorial expression in (6.14) vanishes if $\vep=0$ but this
depends on the vanishing of tensors antisymmetric in five indices so (6.14)
nevertheless represents the singular behaviour of $\Gamma^{CCT}$ as a function
of $d$ as a continuous variable.
In order to obtain a well defined distribution as $d\to 4$ the pole
must be subtracted according to the standard lore of dimensional
regularisation. To calculate $R$ we follow a similar approach to that
followed in section 4 and introduce a factor
$(x-y)^{2\omega_3}(x-z)^{2 \omega_2} (y-z)^{2 \omega_1}$ on the l.h.s. of
(6.14). On the r.h.s. the pole is displaced to 
$\vep +2 \sum_\alpha \omega_\alpha$ and $R$ is modified to $R(\omega)$
although the structure of the residue is unchanged. If we contract indices
then the form of the singularity reduces to
$$ \eqalign {
(x-y)^{2\omega_3}(x-z)^{2 \omega_2} (y-z)^{2 \omega_1} &
\Gamma^{CCT}(x,y,z)_{\mu\ep\eta\rho,\nu\ep\eta\rho,\mu\nu} \cr
& \quad \sim {\R(\omega) \over \vep +2 \sum_\alpha \omega_\alpha}\, 
S_d{}^{\! 2} \de^d(x-z) \de^d(y-z) \, , \cr}
\eqno (6.15) $$
where, inserting (6.2,3), this can now be represented as a sum of terms to which
(4.14) can be applied to determine $\R(\omega)$. It is then clear that
$$
R(\omega) = {\R(\omega) \over f(2\sum_\alpha \omega_\alpha)}
\, ,
\eqno (6.16) $$
where $f(d-4)$ denotes the result of contraction of the tensor in (6.14)
and is given explicitly by (A.2), for $x\to 0$ $f(x) \sim {15\over 4} x$. 
Subject to using the  condition $U=0$ from (6.13) to eliminate one coefficient,
so that for example $I={4\over 3}(4B-2E-G)+4D-2F-2H$, it is straightforward to
take the limit $\omega \to 0$ and obtain for the residue in (6.14)
$$ \eqalign {
R ={}& -{\ts {1\over 32\times 9}}\big ( 2(4B+2E+G+D) + 4( C+D) + 
3 (C'+2D ) - \half F + 7 H + 5 I  \big )  \cr 
={}& {\ts {1\over 16\times 90}} ( 13\A - 2\B - 40\C ) \, , \cr}
\eqno (6.17) $$
where the second line follows from (6.7) although,
since $9 \A - \B - 10\C= 0$ as a result of $C_T=0$, the final expression is
not unique.

Although subtracting the pole in $\vep$ given by (6.14,17) is sufficient to
ensure a regularised expression for $\Gamma^{CCT}$ it is not appropriate when
applied to (6.1) since the symmetry requirements on $\Gamma^{TTT}$ will not
be satisfied. In order to verify that it is sufficient to use the
explicitly symmetric counterterms subtracted in (5.13) we use the result,
with $G$ defined in (5.15),
$$ \eqalign {
{\de\over \de g^{\alpha\beta}(z)} \int \! \d^d x & \,  \sqrt g G = 
{- 15} \, R^{\gamma\de}{}_{[\gamma\de}(z)  R^{\chi\omega}{}_{\chi\omega} (z)
g_{\alpha]\beta}(z) = H_{\alpha\beta}(z) + \vep \, X_{\alpha\beta}(z)\, , \cr
H_{\alpha\beta} = {}& - 15 \, C^{\gamma\de}{}_{[\gamma\de} 
C^{\chi\omega}{}_{\chi\omega} g_{\alpha]\beta} =
2\big ( C^{\gamma\de\chi}{}_{\alpha} C_{\gamma\de\chi\beta} - \quar
g_{\alpha\beta}\, C^{\gamma\de\chi\omega}C_{\gamma\de\chi\omega} \big )\, , \cr
X_{\alpha\beta} = {}& {4\over d-2}\, R^{\gamma\delta} C_{\gamma\alpha\de\beta}
+ 2 {d-3\over(d-2)^2} \big(2 R^\ga {}_{\!\alpha}R_{\ga\beta} - 
g_{\alpha\beta} R^{\ga\de}R_{\ga\de} \big ) \cr
& - \half {d-3\over(d-2)^2(d-1)} \big ( 4d R R_{\alpha\beta} - (d+2)
g_{\alpha\beta} R^2 \big ) \, , \cr}
\eqno (6.18) $$
where the Weyl tensor $C_{\mu\si\rho\nu}$ is defined in (A.3) and the explicit 
form for $ X_{\alpha\beta}$ is unimportant here save for the overall factor
of $\vep$. The variation of the integral must vanish when $d=4$ since it is
then a topological invariant, $H_{\alpha\beta}$ is zero if $d=4$ due to the 
vanishing of antisymmetric five index tensors. Using (A.4) for the form of
the Weyl tensor in an expansion to first order about flat space
$$ \eqalign { \!\!\!\!\!\!\!
{\de\over \de g^{\mu\nu}(x)}{\de\over \de g^{\si\rho}(y)} H_{\alpha\beta}(z)
\Big |_{\rm{flat\ space}}  \!\!\!
& = 16\, \pr^x_{\,\kappa}\pr^x_{\,\lambda} \pr^y_{\,\ep}
\pr^y_{\,\eta}\de^d(x-z)\de^d(y-z) \cr
&\ \times \big (\E^C{}_{\!\!\mu\kappa\lambda\nu,(\alpha|\tau\chi\omega}
\E^C{}_{\!\!\si\ep\eta\rho,|\beta)\tau\chi\omega} - \quar \de_{\alpha\beta}\,
\E^C{}_{\!\!\mu\kappa\lambda\nu,\si\ep\eta\rho} \big ) \, . \cr}
\eqno (6.19) $$
To $\rO(\vep)$ the tensor appearing in (6.19) is identical with 
$\E^C{}_{\!\!\mu\kappa\lambda\nu,\alpha'\tau\chi\omega}
\E^C{}_{\!\!\si\ep\eta\rho,\beta'\tau\chi\omega} 
\E^T{}_{\!\!\alpha'\beta',\alpha\beta}$ which appears in (6.14)
(the difference arises from the $1/d$ term in $\E^T$) so that from (6.1,14)
we may now define a symmetric regularised expression for $d\approx 4$ by
$$
\Ga^{TTT}_{\, \mu\nu,\si\rho,\alpha\beta}(x,y,z)_C
= \pr^x_{\,\kappa}\pr^x_{\,\lambda} \pr^y_{\,\ep}\pr^y_{\,\eta}
\Gamma^{CCT}_{\mu\kappa\lambda\nu,\si\ep\eta\rho,\alpha\beta}(x,y,z) -
\mu^{-\vep}{R\over \vep} \, {S_d{}^{\! 2}\over 16}
D^G_{\, \mu\nu,\si\rho,\alpha\beta}(x,y,z) \, ,
\eqno (6.20) $$
where $D^G$ is defined in (5.14) and is given explicitly by
$$ \eqalign { \!\!
D^G_{\, \mu\nu,\si\rho,\alpha\beta}(x,y,z &) \cr
=  -30 \big\{& \E^{(5)}{}_{\!\!\mu\si\alpha\kappa\ep,\nu\rho\beta\lambda\eta}
\pr^x{}_{\!\kappa}\pr^x{}_{\!\lambda} \pr^y{}_{\!\ep}\pr^y{}_{\!\eta}
\de^d(x-z)\de^d(y-z)+ \mu \leftrightarrow \nu, \si \leftrightarrow \rho \big \}
\, , \cr}
\eqno (6.21) $$
with $\E^{(5)}{}_{\!\!\mu\si\alpha\kappa\ep,\nu\rho\beta\lambda\eta}$ denoting
the projector onto totally antisymmetric five index tensors. From its
definition $D^G_{\, \mu\nu,\si\rho,\alpha\beta}(x,y,z)$ is manifestly
symmetric and the results (6.19) show that subtraction of the counterterm
in (6.21) is equivalent to subtracting the $\vep$ pole in (6.14). It is also
easy to see that $\pr^x_{\, \mu} D^G_{\, \mu\nu,\si\rho,\alpha\beta}(x,y,z)
=0$ so that the counterterm in (6.21) preserves the Ward identity
$\pr^x_{\, \mu} \Ga^{TTT}_{\, \mu\nu,\si\rho,\alpha\beta}(x,y,z)_C =0$.
However the counterterm generates an anomaly in the trace since
$$
D^G_{\,\mu\nu,\si\rho,\alpha\alpha}(x,y,z) = \half \vep \, 
\A^G_{\,\mu\nu,\si\rho}(x-z,y-z) \, ,
\eqno (6.22) $$
where
$$ \eqalign{
\A^G_{\,\mu\nu,\si\rho}(x-z,& y-z) =  
{\de^2\over g^{\mu\nu}(x) g^{\si\rho}(y)}  G(z)  \Big |_{\rm{flat\ space}} \cr
= {}& - 24 \big\{  \E^{(4)}{}_{\!\!\mu\si\kappa\ep,\nu\rho\lambda\eta}
\pr^x{}_{\!\kappa}\pr^x{}_{\!\lambda} \pr^y{}_{\!\ep}\pr^y{}_{\!\eta}
\de^d(x-z)\de^d(y-z) + \si \leftrightarrow \rho \big \} \, . \cr}
\eqno (6.23) $$
Hence for $d=4$ we have the anomaly   
$$
\Ga^{TTT}_{\, \mu\nu,\si\rho,\alpha\alpha}(x,y,z)_C
= - {\ts{1\over 8}}\pi^4 R \, \A^G_{\,\mu\nu,\si\rho}(x-z,y-z) \, ,
\eqno (6.24) $$
where in $\A^G$ we can take 
$24 \E^{(4)}{}_{\!\!\mu\si\kappa\ep,\nu\rho\lambda\eta} \to 
\ep_{\mu\si\kappa\ep} \ep_{\nu\rho\lambda\eta}$.
Comparing the result with (1.9), since in this special case $C_T$ and hence
$\beta_a$ are zero, it is easy to see that we must have
$$
\beta_b = {\ts{1\over 32}}\pi^4 R = 
{\ts {1\over 512\times 90}}\pi^4 ( 13\A - 2\B - 40\C ) \, .
\eqno (6.25) $$

The above analysis based on the expression (6.1) allows the topological
anomaly term to be derived in four dimensions. It is clearly possible to
assume more symmetrically
$$
\Ga^{TTT}_{\, \mu\nu,\si\rho,\alpha\beta}(x,y,z)_C
= \pr^x_{\,\kappa}\pr^x_{\,\lambda} \pr^y_{\,\ep}\pr^y_{\,\eta}
\pr^z_{\,\gamma}\pr^z_{\,\delta}
\Gamma^{CCC}_{\mu\kappa\lambda\nu,\si\ep\eta\rho,\alpha\gamma\delta\beta}
(x,y,z) \, ,
\eqno (6.26) $$
with $\Gamma^{CCC}_{I,J,K}(x,y,z)$, where $I,J,K$ each denote sets of
indices with the symmetry (1.4), an integrable function even in four dimensions.
Assuming manifest conformal invariance we can write
$$
\Gamma^{CCC}_{I,J,K}(x,y,z) = {\I^C{}_{\!\!I,I'}(x-z)\I^C{}_{\!\! J,J'}(y-z)
\over \big ((x-z)^2(y-z)^2\big )^{d-2}}\,
t^{CCC}_{\, I',J', K}(Z) \, .
\eqno (6.27) $$
In this case we follow the treatment in section 2 to obtain a particular
totally symmetric form for $\Gamma^{CCC}_{I,J,K}(x,y,z)$ by writing, as in
(2.25),
$$
t^{CCC}_{\, I,J,K}(Z) = \I^C{}_{\!\! K,K'}(Z) \big ( \X_1 \, d^1{}_{\! IJK'}
+ \X_2 \, d^2{}_{\! IJK'} \big ) {1\over (Z^2)^{\hh (d-2)}} \, ,
\eqno (6.28) $$
where $d^i{}_{\! IJK}$ are the two possible symmetric invariant tensors which
may be given by
$$ \eqalign {
d^1{}_{\! IJK} = \E^C{}_{\!\! I,\ep\eta\kappa\lambda} \,
\E^C{}_{\!\! J,\kappa\lambda\chi\omega}\,
\E^C{}_{\!\! K,\chi\omega\ep\eta} \, , \cr
d^2{}_{\! IJK} = \E^C{}_{\!\! I,\ep\eta\kappa\lambda} \, 
\E^C{}_{\!\! J,\lambda\chi\eta\omega}\,
\E^C{}_{\!\! K,\omega\kappa\chi\ep} \, . \cr}
\eqno (6.29) $$
These results then show that, for arbitrary $d$, that there are two linearly
independent symmetric 
forms for the conformally covariant energy momentum tensor three point function
in which Ward identities and the traceless conditions for the energy
momentum tensor are trivially satisfied. This is in accord with the
result, as in section 5, that there are three in general but that there are
the non trivial Ward identities (2.34,35). However when
$d=4$ the two expressions given by (6.27,28,29) are no longer independent
since
$$
d^1{}_{\! IJK} = 4 d^2{}_{\! IJK} \, ,
\eqno (6.30) $$
which can be shown using the vanishing of antisymmetric five index tensors.
Such a relation must hold otherwise the possibility of a second anomalous
term in the four dimensional trace identity (1.9) would be ruled out.
These formulae can be related to previous results since from (6.26,27) we
may evaluate the $z$ derivatives to give $\Gamma^{CCT}$ as defined by (6.1,2)
together with (6.8),
$$
{\tilde t}^{CCC}_{I,J,K}(X) = \I^C{}_{\!\! J,J'}(X) t^{CCT}_{I,J',K}(-X) \, ,
\qquad {\tilde t}^{CCT}_{I,J,\alpha\beta}(X) = \pr^X_{\, \gamma}
\pr^X_{\, \delta} {\tilde t}^{CCC}_{I,J,\alpha\gamma\delta\beta}(X)\, .
\eqno (6.31) $$
Hence we may determine in terms of $\X_1,\X_2$ the coefficients 
$A,B,C,C',D,E,G,F,H,I$ in (6.3). For general $d$ the results are again not
very succinct so they are given in appendix C. If $d=4$ the linear
combinations in (6.5) depend only on $4\X_1 + \X_2$, as a consequence of the
relation (6.30), and we may then obtain by using (6.7) a unique result for
the three point function which is expressible in the form (6.26)
$$
\A = {\ts {8\over 9}}(4\X_1 + \X_2) \, , \qquad
\B = {\ts {92\over 9}}(4\X_1 + \X_2) \, , \qquad
\C = - {\ts {2\over 9}}(4\X_1 + \X_2) \, .
\eqno (6.32) $$
A consistency check is that $C_T$, as given by (5.12) for $d=4$, and
$\beta_b$ in (6.25) are both zero as required since, as mentioned earlier,
the Ward identities are trivial when the three point function is represented
as in (6.26).
\bigskip
\leftline{\bigbf 7 Effective Actions}
\medskip
In discussing the energy momentum tensor it is natural to introduce an
effective action $W(g)$ whose variational derivatives with respect to
a metric $g^{\mu\nu}$ may be taken as defining correlation functions of
$T_{\mu\nu}$ on a general curved background. Assuming diffeomorphism
invariance is preserved, so that $W(g)$ is a scalar, then Ward identities
may be summarised just by the covariant equation (1.6). If the underlying
quantum field theory is conformally invariant on flat space then $W(g)$
is also invariant under local Weyl transformations of the metric,
$\de_\si g^{\mu\nu} = 2\si g^{\mu\nu}$, apart from the anomalies arising
from $ g^{\mu\nu}\l T_{\mu\nu} \r_g$ in two and four dimensions.

As is well known in two dimensions the trace anomaly may be integrated
uniquely to give \poly\
$$ \eqalign{
W(g) = {}& - {c\over 96\pi} \int \! \! \int \! \d^2 x \d^2 x' \,
\sqrt{g(x)}R(x) G_\Delta (x,x') \sqrt{g(x')}R(x') \, , \cr
& \Delta_x G_\Delta (x,x') = \de^2(x-x') \, , \quad \Delta = - \sqrt g
\nab^2 \, ,\cr}
\eqno (7.1) $$
since under a conformal variation $\de_\si \Delta = 0$ and
$\de_\si( \sqrt g R) = - 2\Delta \si$.
Although $W(g)$ clearly vanishes on flat space it gives a non zero result
for two or more functional derivatives before taking $g_{\mu\nu}=\de_{\mu\nu}$.
To calculate the two point function on flat space arising from $W(g)$ we may 
use $\de_g(\sqrt g R )
= - \sqrt g(\nab_\mu\nab_\nu - g_{\mu\nu} \nab^2) \de g^{\mu\nu}$ and 
then since $G_\Delta(x,x')|_{\rm flat\ space} = {- \ln \mu^2 (x-x')^2/4\pi}$
it is easy to see that
$$
\l T_{\mu\nu} (x) T_{\si\rho}(y) \r = -{c\over 48\pi^2}S^x {}_{\!\! \mu\nu}
S^y {}_{\!\! \si\rho} \ln (x-y)^2 \, , \quad S_{\mu\nu}
= \pr_\mu\pr_\nu - \de_{\mu\nu} \pr^2 \, , 
\eqno (7.2) $$
and hence with complex coordinates $z=x_1+ix_2$ and $T(z) = - 2\pi T_{zz}(x)$
we recover the standard two dimensional conformal field theory result
$$
\l T(z_1) T(z_2) \r = {c\over 2} \, {1\over (z_1-z_2)^4} \, .
\eqno (7.3) $$
The corresponding three point function on flat space, for non-coincident points,
is also easily obtained, using as well $\de_g (\sqrt g \nab^2)= \pr_\mu \sqrt g
( \de g^{\mu\nu} - \half g^{\mu\nu} g_{\si\rho}\de g^{\si\rho} ) \pr_\nu$ in
order to calculate the variation of $G_\Delta$,
$$ \eqalign{
\l T(z_1)& T(z_2) T_(z_3) \r  \cr
= {}& - {c\over 3} \Big \{
{1\over (z_1-z_3)^3 (z_2 -z_3)^3} + {1\over (z_2-z_1)^3 (z_3 -z_1)^3}
+ {1\over (z_1-z_2)^3 (z_3 -z_2)^3} \Big \} \cr
= {}& { c\over (z_1-z_3)^2 (z_2 -z_3)^2 (z_1-z_2)^2} \, , \cr}
\eqno (7.4) $$
which is again in accord with expectation.

In four dimensions there is no possibility of finding any analogous general
result although Barvinsky {\it et al} \barv\ 
have explored in detail approximations
based on an expansion in the curvature. Here we examine a proposal for
part of the effective action due to Riegert \riegert\footnote{*}{For further
discussions and also subsequent references see \odintsov.}
which exactly reproduces
the Euler density term $G$ in the curved space trace anomaly (1.7). This result
is similar in form to the two dimensional expression (7.1)
$$ \eqalign{
W(g)_{\rm Riegert} = {}& {- {\beta_b\over 8}}\int \!\! \int \! \d^4 x \d^4 x' \,
\G(x) G^{\rm R}(x,x')  \G(x')
+ {\beta_b\over 18} \int  \! \d^4 x \, \sqrt g R^2 \, , \cr
& \G = \sqrt g ( G - {\ts {2\over 3}} \nab^2 R) \, , \quad
\Delta^{\rm R} G^{\rm R}(x,x') = \de^4(x-x') \, , \cr
\Delta^{\rm R} & = \sqrt g
\nab^2 \nab^2 + \pr_\mu H^{\mu\nu}\pr_\nu \, , \quad H^{\mu\nu}
= 2\sqrt g ( R^{\mu\nu} - {\ts{1\over 3}}g^{\mu\nu}R ) \, .\cr}
\eqno (7.5) $$
Under a conformal variation $\de_\si g^{\mu\nu} = 2\si g^{\mu\nu}$ then
$\de_\si \G = - 4 \Delta^{\rm R} \si, \, \de_\si 
\Delta^{\rm R} = 0$\footnote{${}^\dagger$}
{The fourth order operator $\Delta^{\rm R}$ which is invariant under
local rescalings of the metric was also found independently by Paneitz and
Eastwood and Singer \pan.}  and also 
$\de_\si (\sqrt g R^2) = 12 \sqrt g R \nab^2 \si$ which is sufficient to show
that $\de_\si W(g)_{\rm Riegert} = \beta_b \int \! \d^4 x \sqrt g \si G$.

Just as in two dimensions we may obtain correlation functions involving the
energy momentum tensor by functional differentiation and then restricting
to flat space. It is straightforward to see that the second derivative
of (7.5) restricted to flat space gives zero. The essential result
which is used to calculate the variation of the metric is
$$ \eqalign {
\de_g \bigg (& \int \!\! \int \! \d^4 x \d^4 y  \sqrt{g(x)}\sqrt{g(y)}\,
\nab^2 X(x) G^{\rm R}(x,y) \nab^2 X(y) - \int  \! \d^4 x \, \sqrt g X^2 \bigg )
\bigg |_{\rm flat \ space} \cr
& = \int \!\! \int \!\! \int \! \d^4 x \d^4 y \d^4 z \, \de g^{\si\rho}(z)
\D^z{}_{\! \si\rho,\ep\eta} \pr_\ep G_0(z-x)X(x) \, \pr_\eta G_0(z-y)
X(y) \, , \cr}
\eqno (7.6) $$
where $G_0$ is the flat space restriction of $-\nab^2 G^{\rm R}$,
$$
G_0(x) = {1\over 4\pi^2 x^2} \, ,
\eqno (7.7) $$
and $\D_{\si\rho,\ep\eta}$ is a differential operator defined by
$$ \eqalign {
& \int  \! \d^4 x \, X_{\ep\eta} \de_g H^{\ep\eta}\Big |_{\rm flat\ space} =
- \int  \! \d^4 x \, \de g^{\si\rho} \D_{\si\rho,\ep\eta} X_{\ep\eta} \, , \cr
 \D_{\si\rho,\ep\eta} X_{\ep\eta} =  -\pr^2 & X_{\si\rho} - \de_{\si\rho}
\pr_\ep\pr_\eta X_{\ep\eta} + \pr_\si \pr_\ep  X_{\ep\rho} + \pr_\rho \pr_\eta
X_{\si\eta} + {\ts{2\over 3}} (\de_{\si\rho} \pr^2 - \pr_\si\pr_\rho )
X_{\ep\ep} \, . \cr}
\eqno (7.8) $$
Using these results we obtain an expression for the energy momentum tensor
three point function, at non coincident points,
 on flat space corresponding to the effective action
(7.5)
$$ \eqalign {
\Ga^{TTT}_{\, \mu\nu,\si\rho,\alpha\beta}(x,y,z)_{\rm Riegert}^{\vphantom g} =
- {\ts{8\over 9}} \beta_b \Big \{& \D^z{}_{\! \alpha\beta,\gamma\de}
\big ( S^x{}_{\!\mu\nu} G_0(x-z) {\overleftarrow \pr}{}^{\! z}{}_{\!\! (\gamma}
\pr^z{}_{\! \delta)} G_0(z-y) {\overleftarrow S}{}^{\! y}
{}_{\!\! \si\rho}\big ) \cr
& + \hbox{cyclic permutations} \Big \} \, . \cr}
\eqno (7.9) $$
The general result is not very transparent but it is not difficult to find
for $s=x-y\to 0$
$$ \eqalign {
\Ga^{TTT}_{\, \mu\nu,\si\rho,\alpha\beta}(& x,y,z)_{\rm Riegert}^{\vphantom g} 
\sim {\ts{8\over 27}} \beta_b \, I_{\mu\nu\si\rho\gamma\de}(s) \,
\pr_\gamma\pr_\delta G_0(y-z) {\overleftarrow S}{}^{\! z}
{}_{\!\! \alpha\beta} \, , \cr
I_{\mu\nu\si\rho\gamma\de}(s) = {}& \big ( 4 \E^T{}_{\!\!\si\rho,\lambda
(\gamma} \pr_{\de)} \pr_\lambda S_{\mu\nu} + 4\E^T{}_{\!\!\mu\nu,\lambda
(\gamma} \pr_{\de)}\pr_\lambda  S_{\si\rho} - s_{(\gamma}\pr_{\delta)}
S_{\mu\nu} S_{\si\rho} \big ) G_0(s) \, . \cr}
\eqno (7.10) $$
Since $I_{\mu\nu\si\rho\gamma\de}(s)= \rO(s^{-6})$ this result is incompatible
with what would be expected from a conformally covariant three point
function, where according to the expressions obtained earlier the leading
singularity should be $\rO(s^{-4})$, reflecting the contribution of the
energy momentum tensor itself in the operator product expansion of two
energy momentum tensors. In consequence the Riegert effective
action, given by (7.5), is also in disagreement with the short distance
behaviour expected for free fields.

The failure of the Riegert action to lead to results on flat space in agreement
with conformal invariance is a consequence its large distance behaviour. From
diffeomorphism invariance and also invariance under local rescalings
of the metric, up to the standard anomalies, the gravitational effective
action is constrained by
$$ \eqalign { 
\int \! \d^4 x \, \big ( \L_v & g^{\mu\nu} +  2\si g^{\mu\nu} \big )
{\de \over \de g^{\mu\nu}} \, W(g) = \int \! \d^4 x \sqrt g \, \si 
\big (\beta_a F + \beta_b G \big ) \, , \cr   
\L_v g^{\mu\nu} = {}& v^\lambda \pr_\lambda g^{\mu\nu} - \pr_\lambda v^\mu
g^{\lambda\nu} - \pr_\lambda v^\nu g^{\mu\lambda} = - \nab^\mu v^\nu
- \nab^\nu v^\mu \, . \cr}
\eqno (7.11) $$
By considering functional derivatives and then restricting to flat space,
and assuming $(\L_v g^{\mu\nu} +  2\si g^{\mu\nu})|_{\rm flat \ space}=0$
which is identical with the conformal Killing equation in (2.2), it is
possible to derive the conformal invariance identities such as (2.36) for
the two point function. However from the Riegert action (7.5), for
asymptotically flat spaces, it is not difficult to see that in general
$\l T_{\mu\nu}(x) \r_{g, \, {\rm Riegert}} = \rO( |x|^{-5})$ for $|x|\to
\infty$ since the leading term involves $S_{\mu\nu}\pr^2 G^{\rm R}(x,x')
{\overleftarrow \pr}{}^\prime{}_{\!\!\alpha}$.
With this asymptotic behaviour it is necessary to restrict $v^\lambda (x)$
to be $\rO(|x|)$ for large $|x|$ in order to avoid surface terms in
deducing (7.11) from $\nab^\mu \l T_{\mu\nu} \r_g =0$. Hence there is no longer
any identity corresponding to special conformal transformations, involving
$b_\mu$ in (2.3), which play the essential role in restricting the form
of two and three point functions on flat space. In two dimensions the effective
action in (7.1) gives
$\l T_{\mu\nu}(x) \r_{g}  = \rO(|x|^{-4})$ which allows conformal invariance
identities to be derived for $\de z = \rO(z^2)$. This is sufficient for 
invariance under the conformal group $O(3,1)$ which is in accord with the
results
(7.3) and (7.4). The action in (7.5) may also be extended to generate the
other terms in the general curved space trace anomaly in (1.7) but similar
difficulties arise in these cases as well and the effective action is also
invariant under constant scale transformations of the metric contrary to the
required behaviour under the renormalisation group. In consequence the Riegert
form for the effective action is not compatible with the form expected from
conventional quantum field theories.

In general the form of the effective action compatible with the trace 
anomaly (1.7) can be written to quadratic order in the curvatures as
$$ \eqalign{
W(g,A)^{(2)} = {}& \half \beta_a \int \! \d^4 x \sqrt g \, C^{\mu\si\rho\nu} 
\Big ( \ln \big ( - \nab^2 /\hmu^2 \big ) - 1 \Big )C_{\mu\si\rho\nu}\cr 
& - \half \kappa  \int \! \d^4 x \sqrt g \, F^{\mu\nu} \Big (
\ln \big ( - \nab^2 /\hmu^2 \big ) - 1 \Big ) F_{\mu\nu} \, , \cr}
\eqno (7.12) $$
where $\hmu =  2e^{-\gamma}\mu $. For constant rescalings of the metric
this generates exactly the result expected from (1.7) for
$\int \! \d^4 x \sqrt g \, g^{\mu\nu} \l T_{\mu\nu} \r_{g,A}$.
The expression (7.12) may also be seen to exactly generate
the flat space results for the energy momentum tensor and vector
current two point functions  given by (2.38), with $C_T$ given by (2.42), 
and (2.17), with $C_V$ given by (2.46). In general the $\ln(-\nab^2)$ 
dependence in (7.12) ensures that $\l T_{\mu\nu} \r$ and $\l V^\mu \r $ have
suitable behaviour at large distances, contrary to what was found in the
Riegert case above. Beyond leading order very lengthy expressions for
the contributions to $W(g,A)$ cubic in the curvature have been given 
but it is not clear at present what minimal
form to take for these so as to generate just the conformally
covariant energy momentum tensor three point functions on flat space which
have been discussed earlier.

As a simple illustration of an effective action which generates a trace
anomaly on curved space and is compatible with conformal invariance when
reduced to flat space we consider $W(g,J)$, where $J(x)$ is a source
coupled to a dimension two operator $\O$. The local trace anomaly is
then $g^{\mu\nu}\l T_{\mu\nu} \r_{g,J} = p\, \half J^2$. To obtain $W$
we define the four dimensional conformal operator $\Delta$ and its Green
function by
$$
\Delta = \sqrt g \big ({-\nab^2} + {\ts{1\over 6}} R \big ) \, , \quad
\Delta_x G_\Delta (x,x') = \de^4(x-x') \, .
\eqno (7.13) $$
Clearly $G_\Delta (x,x')|_{\rm flat \ space} = G_0(x-x')$ as in (7.7).
The contribution to the effective action involving $J$ may now be taken as
$$
W(g,J) = 4\pi^2 p  \int \! \! \int \! \d^4 x \d^4 x' \,
\sqrt{g(x)}J(x) \R \big ( G_\Delta (x,x')^2\big ) \sqrt{g(x')}J(x') \, ,
\eqno (7.14) $$
with the regularised product,
$$
 \R \big ( G_\Delta (x,x')^2\big ) = \Big ( \mu^{2\omega} 
G_\Delta (x,x')^{2-\omega} - {1\over 16\pi^2 \omega} \de^4(x,x') \Big )
\Big |_{\omega \to 0}\, , 
\eqno (7.15) $$
defined similarly to (2.39), with $\de^4(x,x') = \de^4(x-x')/\sqrt{g(x)}$.
Using for the variation under rescaling of the metric
$\de_\si G_\Delta (x,x') = \big (\si(x)+\si(x')\big ) G_\Delta (x,x')$,
which follows from $\de_\si \Delta = -\si \Delta - \Delta \si$, 
it is easy to see that
$$
\de_\si \R \big ( G_\Delta (x,x')^2\big ) - 2\big (\si(x)+\si(x')\big )
\R \big ( G_\Delta (x,x')^2\big ) = -{1\over 8\pi^2} \, \si(x) \de^4(x,x') \, ,
\eqno (7.16) $$
and hence (7.14) generates the expected form for $\de_\si W(g,J)$, taking
$\de_\si J = 2\si J$. From (7.14) it is trivial to derive the form of
the two point function on flat space giving
$$
\l \O(x) \O(y) \r = {C_\O \over (x-y)^4} \, , \qquad C_\O = {p\over 2\pi^2}\, .
\eqno (7.17) $$
More significant is the calculation of the three point function involving the
energy momentum tensor in which it is necessary to vary the metric. Using
$$
\de_g G_\Delta (x,y) = - \int \! \d^4 z \, G_\Delta (x,z) \de_g \Delta_z
G_\Delta (z,y) \, ,
\eqno (7.18) $$
we may obtain from the definition of $\Delta$ in (7.13)
$$ \eqalign {
{\de\over \de g^{\alpha\beta}(z)} G_\Delta (x,y) \Big |_{\rm flat \ space}
= {}&{1\over 12\pi^4}\, {(x-y)^2\over(x-z)^4(y-z)^4} \Big ( {Z_\alpha Z_\beta
\over Z^2} - {\quar} \de_{\alpha\beta} \Big ) \cr
& + {1\over 32\pi^2} \, {1\over (x-y)^2}
\de_{\alpha\beta} \big ( \de^4(x-z) + \de^4(y-z) \big ) \, , \cr}
\eqno (7.19) $$
and hence, at non coincident points,
$$ \eqalign {
\l \O(x) \O(y) T_{\alpha\beta}(z) \r ={}& - 2{\de^3 \over 
\de J(x)\de J(y)\de g^{\alpha\beta}(z)} W(g,J) \Big |_{\rm flat \ space} \cr
={}&  - {4\over 3\pi^2} \, {C_\O\over 
(x-z)^4(y-z)^4}\Big ( {Z_\alpha Z_\beta\over Z^2} - {\quar} \de_{\alpha\beta} 
\Big ) \, . \cr}
\eqno (7.20) $$
This expression is clearly compatible with conformal invariance, unlike the
results obtained from the Riegert action earlier.

In order to construct  a wider class of effective actions we consider now
a second order differential operator $\Delta^F$ acting on antisymmetric tensor
fields, or 2-forms, $F_{\mu\nu}$ which has simple transformation properties
under local rescalings of the metric analogous to the operator $\Delta$ acting
on scalar fields and defined in (7.13)\footnote{*}{This
operator is a special case of a wider class of conformally
covariant differential operators \branson.}. $\Delta^F$ is defined by
$$ \eqalign {
\Delta^F  F_{\mu\nu} = {}&\sqrt g \big ( [(\delta \d - \d \delta)F]_{\mu\nu}
+ R_{\mu}{}^{\!\lambda} F_{\lambda\nu} + R_{\nu}{}^{\!\lambda} F_{\mu\lambda}
- \half R F_{\mu\nu} + t \, C_{\mu\nu}{}^{\si\rho} F_{\si\rho} \big ) \, , \cr
(\d \delta F)_{\mu\nu} ={}& -2\pr_{[\mu} \Big ( {1\over \sqrt g} g_{\nu]\omega}
\pr_\lambda \big ( \sqrt g g^{\lambda \si} g^{\omega\rho} F_{\si\rho}\big )
\Big ) = - 2\nab_{[\mu} \nab^\lambda F_{\lambda|\nu]} \, , \cr
(\delta \d F)_{\mu\nu} = {}& - {1\over \sqrt g} g_{\mu\gamma} g_{\nu\delta}
\pr_\lambda \big ( \sqrt g g^{\lambda\tau}g^{\gamma\si}g^{\de\rho}
3\pr_{[\tau} F_{\si\rho]} \big ) = - 3 \nab^\lambda \nab_{[\lambda}
F_{\mu\nu]} \, , \cr}
\eqno (7.21) $$
where $t$ is an arbitrary parameter and we have used $\d$ to denote the 
exterior derivative and $\delta$ its adjoint. It is not difficult to see that
$$
\de_{\si} \Delta^F = - 3\si \Delta^F + \Delta^F \si \, , 
\eqno (7.22) $$
which is the essential property for our purposes. From (7.21) $\Delta^F$
may be expressed alternatively by
$$ \eqalign {
\Delta^F F_{\mu\nu} = \sqrt g \big ( &
-\nab^2  F_{\mu\nu} + 2 \nab_\mu \nab^\lambda F_{\lambda \nu} 
- 2 \nab_\nu \nab^\lambda F_{\lambda \mu} \cr
& + R_{\mu}{}^{\!\lambda} F_{\lambda\nu}
- R_{\nu}{}^{\!\lambda} F_{\lambda\mu} - {\ts{1\over 6}} R F_{\mu\nu}
+ (t-1) C_{\mu\nu}{}^{\si\rho} F_{\si\rho} \big ) \, , \cr}
\eqno (7.23) $$
and the basic Green function is defined by
$$
\big ( \Delta^F{}_{\! x} G^F \big ){}_{\mu\nu}^{\ \ \, \si\rho} (x,y)
= \de_\mu^{[\si} \de_\nu^{\rho]} \, \de^4 (x-y ) \, . 
\eqno (7.24) $$
Using (7.22) we may see that
$$
\de_\si G^F{}_{\!\!\! \mu\nu\si\rho}(x,y) = - \big ( \si(x) + \si(y) \big )
G^F{}_{\!\!\! \mu\nu\si\rho}(x,y) \, .
\eqno (7.25) $$
Reducing to flat space, by using Fourier transforms, the Green function is
explicitly given by
$$
G^F{}_{\!\!\! \mu\nu\si\rho}(x,y) \big |_{\rm flat \ space} = -
{1\over 4\pi^2 s^2} \, \I^F{}_{\!\!\mu\nu,\si\rho}(s) \, , \quad s=x-y \, .
\eqno (7.26) $$
Using the analogous result to (7.18) we may also find, for non coincident
points,
$$ \eqalign { \!\!\!\!
{\de\over \de g^{\alpha\beta}(z)} G^F{}_{\! \mu\nu\si\rho} (x,y) 
\Big |_{\rm flat \ space} \!\!\!\!
& = {\I^F{}_{\!\!\mu\nu,\mu'\nu'}(x-z) \I^F{}_{\!\!\si'\rho',\si\rho}(z-y)
\over (4\pi^2)^2 (x-z)^2 (z-y)^2 } \, h_{\mu'\nu',\si'\rho',\alpha\beta}(Z)\cr
{} -{}& 2t \, \E^C{}_{\!\! \alpha\gamma\de\beta,\kappa\lambda\ep\eta}
\pr^z{}_{\!\!\gamma} \pr^z{}_{\!\!\de} \bigg (
{\I^F{}_{\!\!\mu\nu,\kappa\lambda}(x-z) \I^F{}_{\!\!\ep\eta,\si\rho}(z-y)
\over (4\pi^2)^2 (x-z)^2 (z-y)^2 } \bigg ) \, , \cr}
\eqno (7.27) $$
where
$$ \eqalign {
h_{\,\mu\nu,\si\rho,\alpha\beta} (Z) = {}&
16 \, \E^F{}_{\!\! \mu\nu,\lambda\ep} \E^F{}_{\!\! \si\rho,\lambda\eta}
\E^T{}_{\!\! \ep \eta,\alpha\beta} Z^2 \cr
& - 8\, \E^F{}_{\!\! \mu\nu,\lambda\ep} \E^F{}_{\!\! \si\rho,\lambda\eta}
\big ( \E^T{}_{\!\! \ep \kappa,\alpha\beta} Z_\eta +
\E^T{}_{\!\! \eta \kappa,\alpha\beta} Z_\ep \big ) Z_\kappa \cr
& + 4\, \E^F{}_{\!\! \mu\nu,\si\rho} \big ( Z_\alpha Z_\beta 
- \quar \de_{\alpha\beta} Z^2\big ) \, . \cr}
\eqno (7.28) $$

As an application of these results we consider a contribution to the
effective action involving the gauge field $A$ which is of the form
$$
\!\!\!\!  W(g,A)^F\! = -{\ts{1\over 16}} U \! \int \! \! \int \! \d^4 x \d^4 x'
\, \sqrt{g(x)}F^{\mu\nu}(x) G^F{}_{\!\!\!\mu\nu\si\rho}(x,x')
G_\Delta (x,x') \sqrt{g(x')}F^{\si\rho}(x') \, .
\eqno (7.29) $$
It is not difficult to see that no regularisation is necessary in this case
and furthermore, as a consequence of (3.6), there is no contribution to the
two point function of the conserved vector current $V_\mu$ coupled to $A$
on flat space. However (7.29) does imply an expression for the three point
function involving the energy momentum tensor which is given by
$$ \eqalign{
\l V_\mu (x) V_\si(y) T_{\alpha\beta} (z) \r^F = {}& - 2{\de^3 \over
\de A^\mu(x) \de A^\si(y)\de g^{\alpha\beta}(z)} W(g,A)^F 
\Big |_{\rm flat \ space} \cr
= {}& \pr^x{}_{\!\! \nu} \pr^y{}_{\!\!\rho}
\Gamma_{\mu\nu,\si\rho,\alpha\beta}(x,y,z)^F \, , \cr}
\eqno (7.30) $$
where, using the relation
$$
\I^F{}_{\!\!\mu\nu,\si\rho}(x-y) =  \I^F{}_{\!\!\mu\nu,\ep\eta}(x-z)
\I^F{}_{\!\!\ep\eta,\si\rho}(z-y) - 4
\I^F{}_{\!\!\mu\nu,\lambda\ep}(x-z) \I^F{}_{\!\!\lambda\eta,\si\rho}(z-y)
{Z_\ep Z_\eta\over Z^2} \, ,
\eqno (7.31) $$
which follows from (2.9),
$\Gamma_{\mu\nu,\si\rho,\alpha\beta}(x,y,z)^F$ is expressible exactly
in the form (4.2,3), for $d=4$, with
$$
A= \hU \big (16 + {\ts{8\over 3}} t \big ) \, , \
B =- \hU {\ts{16\over 3}} t \, , \ 
C=- \hU \big (8 + {\ts{8\over 3}} t \big ) \, , \ D = \hU \big(
{\ts{8\over 3}} + {\ts{16\over 3}} t \big ) \, , \ E = \hU
{\ts{16\over 3}} \, ,
\eqno (7.32) $$
for $\hU=U/(4\pi^2)^3$. Note that these results satisfy $K=0$, as given in
(4.7), and from (4.12) $I+J= -B-D+\half E =0$ in accord with the Ward 
identities.

For the purely gravitational effective action we may also consider an
analogous contribution to (7.29),
$$
\!\!\!  W(g)^F = {\ts{1\over 64}} V \! \int \! \! \int \! \d^4 x \d^4 x' \,
\sqrt{g(x)}C^{\mu\kappa\lambda\nu}(x) G^F{}_{\!\!\!\mu\kappa\si\ep}(x,x')
G^F{}_{\!\!\! \lambda\nu\eta\rho} (x,x')\sqrt{g(x')}C^{\si\ep\eta\rho}(x') \, .
\eqno (7.33) $$
the singularity as $x\to x'$ again does not require regularisation and, as for
(7.29), $W(g)^F$ does not contribute to the energy momentum tensor two point
function on flat space but there is a corresponding expression for the three 
point function given by
$$ \eqalign { \!\!\!\!\!
\l T_{\mu\nu} (x) T_{\si\rho}(y) T_{\alpha\beta} (z) \r^F \! = {}& -8
{\de^3 \over \de g^{\mu\nu}(x) \de g^{\si\rho}(y) \de g^{\alpha\beta}(z)} 
W(g)^F \Big |_{\rm flat \ space} \cr
= {}& \pr^x{}_{\!\!\kappa}\pr^x{}_{\!\!\lambda} \pr^y{}_{\!\!\ep}
\pr^y{}_{\!\!\eta}
\Gamma_{\mu\kappa\lambda\nu,\si\ep\eta\rho,\alpha\beta}(x,y,z)^F \! + 
\hbox{cyclic permutations} \, . \cr}
\eqno (7.34) $$
$\Gamma_{\mu\kappa\lambda\nu,\si\ep\eta\rho,\alpha\beta}(x,y,z)^F$ may
be represented just as in (6.2,3) for $d=4$ where, by applying (7.27)
and (7.31) again, the coefficients are
$$ \eqalign {
A= {}& -\quar C' = \hV \big ( 32 + {\ts{16\over 3}} t \big ) \, , \quad
B= -\quar G = \hV \big ( 8 + {\ts{32\over 3}} t \big ) \, , \cr
E= {}& -\quar H = \hV \big ( 16 + {\ts{16\over 3}} t \big ) \, , \quad
C =  -\quar F = - \hV {\ts{32\over 3}} t \, , \cr}
\eqno (7.35) $$
for $\hV = V/(4\pi^2)^3$. These results satisfy $T_1 = T_2 = 0$ from (6.11)
and (6.7) now gives
$$
\A = - 4 \times {\ts{32\over 3}} \hV \, , \quad
\B= - 46 \times {\ts{32\over 3}} \hV \, , \quad 
\C = {\ts{32\over 3}} \hV \, .
\eqno (7.36) $$
The ratios are just as in (6.32) as expected since both trace anomalies are
absent for this gravitational effective action.
\bigskip
\leftline{\bigbf 8 Conclusion}
\medskip
In more than two dimensions correlation functions involving the energy
momentum tensor or conserved currents in conformal field theories are not
unique. The results of this paper show that this may be regarded as a
reflection of the freedom in (1.3) or (1.2). When there is no non trivial
Ward identity the expressions can be written so that effectively the energy
momentum tensor or conserved current is exactly given by the trivial form
so that the conservation and vanishing of the energy momentum tensor trace
are automatic. As an illustration we may consider the three point function for
a conserved vector current and two operators $\O_1 , \O_2$ of differing
dimensions $\eta_1 , \eta_2$. The general formula (2.20) gives
$$ \eqalign {
\l V_\mu(x)& \, \O_1^{i} (y) \, \O_2^{j} (z) \r  \cr
& = {1\over (x-z)^{2(d-1)}\,(y-z)^{2\eta_1}} \,
I_{\mu\mu'}(x-z)
D_1^{\, i} {}_{i'} (I(y-z)) \, t_{\mu'}{}^{i'j} (Z) \, , \cr }
\eqno (8.1) $$
where $t_{\mu}{}^{ij} (Z) = \rO(Z^{-(d-1+\eta_1-\eta_2)})$ and from the
conservation equation $\pr_\mu t_{\mu}{}^{ij} (Z) =0$. It is easy to see
that
$$
(\eta_1  - \eta_2 ) t_{\mu}{}^{ij} (Z) = \pr_\nu \big ( Z_\mu t_{\nu}{}^{ij}(Z)
- Z_\nu t_{\mu}{}^{ij} (Z) \big ) \, ,
\eqno (8.2) $$
and hence, for $\eta_1 \ne \eta_2$, we can always write $t_{\mu}{}^{ij} (Z)
= \pr_\nu F_{\mu\nu}{}^{\! ij} (Z)$, $F_{\mu\nu}{}^{\! ij}(Z)=
- F_{\nu\mu}{}^{\!ij}(Z)$, so that (8.1) can be written as
$$ \eqalign {
\l V_\mu & (x) \, \O_1^{i} (y) \, \O_2^{j} (z) \r  \cr
& = \pr^x{}_{\!\! \nu}\bigg({1\over (x-z)^{2(d-2)}\,(y-z)^{2\eta_1}} \,
\I^F{}_{\!\!\mu\nu,\mu'\nu'}(x-z)
D_1^{\, i} {}_{i'} (I(y-z)) \, F_{\mu'\nu'}{}^{\! i'j} (Z)\bigg ) \, . \cr }
\eqno (8.3) $$
A similar discussion applies in the case of the energy momentum tensor where
the corresponding expression for the three point function has the form
$$ \eqalign {
\l T_{\mu\nu}(x)& \, \O_1^{i} (y) \, \O_2^{j} (z) \r  \cr
& = {1\over (x-z)^{2d}\,(y-z)^{2\eta_1}} \,
\I^T{}_{\!\!\mu\nu,\mu'\nu'}(x-z)
D_1^{\, i} {}_{i'} (I(y-z)) \, t_{\mu'\nu'}{}^{i'j} (Z) \, , \cr }
\eqno (8.4) $$
for $ t_{\mu\nu}{}^{ij} (Z) = \rO(Z^{-(d+\eta_1-\eta_2)})$.
In this case, for $\eta_1 \ne \eta_2$, it is possible to write
$ t_{\mu\nu}{}^{ij}(Z) = \pr_\si \pr_\rho C_{\mu\si\rho\nu}{}^{ij}(Z)$ with 
$C_{\mu\si\rho\nu}{}^{ij}(Z)$ satisfying (1.4) and hence, analogous to (8.3),
we may  pull out two derivatives so there are no non trivial Ward identities,
reflecting that the two point function $\l \O_1^{j} (y) \, \O_2^{k} (z) \r
=0$. To demonstrate the existence of $C_{\mu\si\rho\nu}{}^{ij}(Z)$ we consider
the Fourier transform ${\tilde t}_{\mu\nu}{}^{ij}(k)= \rO(k^{(\eta_1-\eta_2)})$.
When $\eta_1 \ne \eta_2$ the Fourier transform is unambiguous and
satisfies $k_\mu {\tilde t}_{\mu\nu}{}^{ij}(k) = 0 , \,
{\tilde t}_{\mu\mu}{}^{ij}(k) = 0 $ (for $\eta_1 = \eta_2$ 
${\tilde t}_{\mu\nu}(k)$ is ambiguous up to a constant $C_{\mu\nu}$ and
the equations become $k_\mu {\tilde t}_{\mu\nu}(k) = k_\mu a_{\mu\nu}, \,
{\tilde t}_{\mu\mu}(k) = b$ with $a_{\mu\nu},\, b$ constrained by Ward
identities as in (2.31)). Subject to these results we may then define
$$
{\tilde C}_{\mu\si\rho\nu}{}^{ij}(k) = - {1\over (k^2)^2}\, {4(d-2)\over d-3}
\, \E^C{}_{\!\!\mu\si\rho\nu,\alpha\gamma\delta\beta}k_\gamma k_\delta \,
{\tilde t}_{\alpha\beta}{}^{ij}(k) \, ,
\eqno (8.5) $$
since from (A.5) it is clear that $-k_\si k_\rho 
{\tilde C}_{\mu\si\rho\nu}{}^{ij}(k) = {\tilde t}_{\mu\nu}{}^{ij}(k)$. 
For some cases, such as when $\O_1 , \O_2$ are scalar fields, there is no
solution for $t_{\mu}{}^{ij} (Z)$ or $t_{\mu\nu}{}^{ij} (Z)$ satisfying the
necessary conditions so that the three point function must vanish (for
$\O_1 , \O_2$ scalars this may
be seen from (8.2) since $t_{\mu}(Z) \propto Z_\mu$).

In a conformal field theory the coefficients of the trace anomaly depending on
the Riemann curvature for a
space background appear to play a significant role whose consequences may
not yet be fully understood. As Deser and Schwimmer \deser\ have made clear
in even dimensions, larger than two, there are two distinct classes of terms
which may contribute to this anomaly. In four
dimensions the term proportional to $\beta_b$ has an apparent topological
significance and, as shown in section 6, in a dimensional regularisation
context is related to an $\rO(\vep/\vep)$ counterterm since it involves
tensorial expressions which vanish identically for $d=4$. From this point of
view it is directly related to the Virasoro central charge $c$ in two 
dimensions. The other terms present in the trace anomaly derive from
counterterms constructed from the Weyl tensor and hence are related to standard
short distance divergences. In four
dimensions there is only one anomaly term formed in this fashion. This has the 
coefficient $\beta_a$ which then determines the scale of the two point function
for the energy momentum tensor.
Both $\beta_a$ and $\beta_b$ are also connected to the corresponding
three point function. It would be interesting to see if there is a way of
projecting out the part proportional to $\beta_b$ which might make feasible
an analysis similar to that of Zamolodchikov \zam\ in two dimensions. A
still unresolved question is whether there is any requirement for $\beta_b$
to be positive. As a conjecture it may be possible to apply the various
suggested positivity
conditions of classical general relativity \haw\ to three point functions
involving the energy momentum tensor but unfortunately the most naive
application of such ideas does not lead to any  direct condition on 
$\beta_b$ \proc. As an illustration of such constraints, for scalar operators
$\O$ in a field theory on 
Euclidean space, if $n_\mu$ is any unit vector we may impose the reflection
positivity condition  $\l \O(\lambda n) T_{nn}(0)
\O(-\lambda n) \r < 0 $ where $T_{nn}= n_\mu n_\nu T_{\mu\nu}$, which
is related to the notion of positive energy density. In the conformal
limit this is equivalent to just positivity of the two point function for the
scalar operator $\O$ itself but without conformal invariance this is a
possible independent condition on the quantum field theory. A separate
consequence of the results obtained here relates to implications for the
gravitational effective action which may describe the back reaction of the
matter fields 
on a curved space background. It is clear from our discussion of the
Riegert action that additional constraints on the fall off of the energy
momentum tensor expectation value at large distances need to be imposed
if the results, when reduced to flat space, are to correspond to standard
field theory results.
\bigskip
\leftline{\bigbf Acknowledgements}
\medskip
We are grateful to Andrei Barvinsky for very helpful correspondence concerning
the Riegert action.
\vfill\eject
\leftline{\bigbf Appendix A}
\medskip
The projector operator $\E^C$ has the explicit form
$$\eqalignno{
\E^C{}_{\!\! \mu\si\rho\nu,\alpha\gamma\delta\beta} = {}&
{\ts {1\over 12}}\big( \de_{\mu\alpha}\de_{\nu\beta}\de_{\si\gamma}\de_{\rho\de}
+ \de_{\mu\de}\de_{\si\beta}\de_{\rho\alpha}\de_{\nu\gamma}
- \mu \leftrightarrow \si, \nu \leftrightarrow \rho \big ) \cr
& + {\ts{1\over 24}}\big( \de_{\mu\alpha}\de_{\nu\gamma}\de_{\rho\de}\de_{\si\beta}
- \mu \leftrightarrow \si, \nu \leftrightarrow \rho, 
\alpha \leftrightarrow \gamma , \de \leftrightarrow \beta  \big ) \cr
& - {1\over d-2} \, {\ts {1\over 8}}\big( 
\de_{\mu\rho}\de_{\alpha\delta}\de_{\si\gamma}\de_{\nu\beta}
+ \de_{\mu\rho}\de_{\alpha\delta}\de_{\si\beta}\de_{\nu\gamma}
- \mu \leftrightarrow \si, \nu \leftrightarrow \rho, 
\alpha \leftrightarrow \gamma , \de \leftrightarrow \beta  \big ) \cr
& + {1\over (d-2)(d-1)}\, {\ts {1\over 2}}\big( 
\de_{\mu\rho} \de_{\nu\si} - \de_{\mu\nu}\de_{\si\rho} \big ) \big (
\de_{\alpha\de}\de_{\beta\gamma} - \de_{\alpha\beta}\de_{\gamma\de} \big ) \, . 
& (A.1) \cr}
$$
The manipulation of such tensors is obviously time consuming and throughout
we made extensive use of FORM \form. The dimension of the space of tensors with
the symmetry properties (1.4) in $d$ dimensions is 
$\E^C{}_{\!\! \mu\si\rho\nu,\mu\si\rho\nu}={1\over 12}d(d+1)(d+2)(d-3)$.
For use in section 6 we may for instance calculate
$$
\E^C{}_{\!\!\mu\ep\eta\rho,\alpha\tau\chi\omega}
\E^C{}_{\!\!\nu\ep\eta\rho,\beta\tau\chi\omega}
\E^T{}_{\!\!\alpha\beta,\mu\nu}
= {(d-1)(d-3)(d-4)(d+4)(d+2)(d+1)\over 96(d-2)} \equiv f(d-4) \, .
\eqno (A.2) $$
The factor $d-4$ is expected since the tensorial expression vanishes
identically when $d=4$.

On a curved space background with a metric $g_{\mu\nu} = \de_{\mu\nu} +
h_{\mu\nu}$ then the Weyl tensor
$$
C_{\mu\si\rho\nu} = R_{\mu\si\rho\nu} - {2\over d-2} \bigl (
g_{\mu[\rho}R_{\nu]\si} - g_{\si[\rho}R_{\nu]\mu} \bigl )
{} + {2\over (d-1)(d-2)}\, g_{\mu[\rho}g_{\nu]\si}  R \, , 
\eqno (A.3) $$
to first order in $h$ can be written as
$$
C_{\mu\si\rho\nu} = 2\, \E^C{}_{\!\! \mu\si\rho\nu,\alpha\gamma\delta\beta}
\, \pr_\gamma \pr_\de h_{\alpha\beta} \, .
\eqno (A.4) $$

A consequence of the definition (A.1), which is useful in the text, is to
consider contraction with a vector $k$,
$$ \eqalign { \!\!\!\!\!\!\!\!\!\!
\E^C{}_{\!\! \mu\si\rho\nu,\alpha\gamma\delta\beta} & k_\si k_\rho 
k_\gamma k_\delta = {d-3\over 4(d-2)}\bigg ( {d-2\over d-1}\, k_\mu k_\nu
k_\alpha k_\beta - \half k^2 \big (k_\mu k_\alpha \de_{\nu\beta}
+ \mu \leftrightarrow \nu, \alpha \leftrightarrow \beta \big ) \cr
+ {1\over d-1} k^2 & \big ( k_\mu k_\nu \de_{\alpha\beta} 
+ k_\alpha k_\beta \de_{\mu\nu} \big ) + \half (k^2)^2 \big ( \de_{\mu\alpha}
\de_{\nu\beta} + \de_{\mu\beta} \de_{\nu\alpha} \big ) -
{1\over d-1} (k^2)^2 \de_{\mu\nu} \de_{\alpha\beta} \bigg ) \, . \cr}
\eqno (A.5) $$
\bigskip
\leftline{\bigbf Appendix B}
\medskip
For general $d$ the trivial free boson and fermion theories give for the
coefficients $\A,\B,\C$ the results \one\
$$\eqalign{
\A ={}& {1\over S_d{}^{\! 3}}\, {d^3\over (d-1)^3}\, n_S \, , \cr
\B ={}& - {1\over S_d{}^{\! 3}} \Big( {(d-2)d^3\over (d-1)^3}\, n_S +
2d^2\,{\tilde n}_F \Big) \, , \cr
\C ={}& - {1\over S_d{}^{\! 3}} \Big( {(d-2)^2d^2 \over 4(d-1)^3}\, n_S +
d^2\,{\tilde n}_F \Big) \, . \cr }
\eqno (B.1) $$
Here ${\tilde n}_F = \quar \tr (1) n_F$, where $\tr$ is the Dirac trace. From
(5.12) we get the standard result
$$
C_T = {1\over S_d{}^{\! 2}} \Big( {d\over d-1}\, n_S + 2d \, {\tilde n}_F 
\Big) \, .
\eqno (B.2) $$
\leftline{\bigbf Appendix C}
\medskip
We here give the results for general $d$ for the coefficients determining
the energy momentum tensor three point function obtained in section 6,
$$ \eqalignno { \!\!\!\!
\A = {}& - {d(d-4)^2\over 2(d-2)} A + 2 \, {d(d-3)\over d-1} B\cr
& + {d\over (d-1)^2(d-2)} \Big \{
\half (d^3-6d^2+9d+4)C + \half ( d^3-7d^2+15d-5)C' \cr
&\qquad \qquad \qquad \qquad + (d-3)(d^2-d+2)  D \Big \} \cr
& + {d\over (d-1)(d-2)} \Big \{ (d^2-7d+14)E - (d-5)G -\half (d-9) H + 2I
\Big \} \cr
& - {d\over 4(d-1)^2(d-2)^2}(3d^3-16d^2+15d+6) F \, , & (C.1a) \cr
\B = {}& - {d(d-4)^2(d-1)\over 4(d-2)} A + 2 \, {d^2(d-3)\over d-1} B
+ {d\over 2(d-1)(d-2)} (3d^3-18d^2+27d+4)E \cr
& + {d\over (d-1)^2(d-2)} \Big \{
- \quar  (d^4-7d^3+11d^2+11d-32)C \cr
&\qquad \qquad \qquad \quad + \quar ( d^4-7d^3+17d^2-17d+14)C' 
+ (d-3)(d^3-3d^2+2d+2)  D \Big \} \cr
& - {d\over 2(d-1)(d-2)^2} \Big \{ (d^3-10d^2+21d-4)G \cr
&\qquad \qquad \qquad \qquad +\half d(d-5)(3d-7) H 
- (3d^2-8d+1) I \Big \} \cr
& + {d\over 8(d-1)^2(d-2)^2}(d^4-7d^3+11d^2+27d-48) F \, , & (C.1b) \cr
\C = {}& - {d(d-4)^2(d-1)\over 8(d-2)} A + {d(d-2)(d-3)\over 2(d-1)} B
+ {d\over 4(d-1)(d-2)} (d^3-10d^2+33d-32)E \cr
& + {d\over 4(d-1)^2(d-2)} \Big \{
+ \half  (d^4-5d^3-d^2+37d-40)C \cr
&\qquad \qquad \qquad \qquad + \half ( d^4-8d^3+21d^2-14d-4)C' 
+ (d-3)(d^3-3d^2+8d-8)  D \Big \} \cr
& - {d\over 4(d-1)(d-2)^2} \Big \{ (d^3-8d^2+19d-16)G \cr
&\qquad \qquad \qquad \qquad +\half (d^3-14d^2+41d-36) H 
- (2d^2-7d+7) I \Big \} \cr
& - {d\over 16(d-1)^2(d-2)^2}(3d^4-15d^3+d^2+63d-60) F \, . & (C.1c) \cr}
$$
For $d=4$ these reduce to the results given in (6.8).

The relation of the expression for $\Ga^{CCC}$ to the general form for
$\Ga^{CCT}$ for arbitrary $d$ is given by the relations
$$ \eqalignno{
A = {}& {- {1\over(d-1)(d-2)}}\big ( d(d^2-d-16)\X_1 - \half
(d^3 - 4d^2 - 7d - 6) \X_2 \big ) \, , \cr
B = {}& {1\over(d-1)(d-2)}\big ( d^2\X_1 + \half
(2d^2 - 9d + 6) \X_2 \big ) \, , \cr
C = {}& {- {1\over(d-1)(d-2)}}\big ( (d^4-3d^3-10d^2+20d+56)\X_1 - 
(d^3 - 11d^2 + 18d + 24) \X_2 \big ) \, , \cr
C' = {}& {- {1\over(d-1)(d-2)}}\big ( 16(d-5)\X_1 -  \half
(d^4 - 11d^3 + 42d^2 -48d-48) \X_2 \big ) \, , \cr
D = {}& {1\over(d-1)(d-2)}\big (4(d^2-3d-6)\X_1 + \quar
(d^4-9d^3+20d^2-12d+64) \X_2 \big ) \, , \cr
E = {}& {1\over(d-1)(d-2)}\big ( (d^3-d^2-28d+12)\X_1 -
(2d^3 -11d^2 +7d - 6) \X_2 \big ) \, , \cr
F = {}& 0 \, ,  \qquad
G = {6 \over(d-2)}\big ( 8 \X_1 +(d^2 - 5d + 2) \X_2 \big ) \, , \cr
I = {}& {- 2H} = 
4{d-4\over(d-2)}\big ( 8 \X_1 +(d^2 - 5d + 2) \X_2 \big ) \, . & (C.2)\cr}
$$
\listrefs
\bye